\documentclass{aa}

\usepackage{graphicx}
\usepackage{verbatim}
\usepackage{color}
\usepackage{threeparttable}
\usepackage{upgreek}
\usepackage{hyperref}
\usepackage{verbatim}
\usepackage{color}
\usepackage{upgreek}
\newcommand{\Msun}{$M_{\odot}$}
\newcommand{\asec}{^{\prime \prime}}
\def\xmm{XMM--{\it Newton}}

\begin{document}

   \title{The plethora of diffuse emission in Abell~2034 as revealed by MeerKAT polarization observations} 

   \subtitle{}

   \author{A. Bonafede
          \inst{1,2}
           \and M. Balboni
          \inst{1,9}
            \and G.W. Pratt\inst{3}
         \and          I. Bartalucci
    \inst{9}
          \and L.~Rudnick
          \inst{8}
           \and C.~J.~Riseley
          \inst{6,7}
         \and C.~Stuardi
          \inst{2}  
          \and B.~Hugo
          \inst{12,11}
          \and{G. Bernardi}
          \inst{2,11,12}
          \and M. Br\"uggen
          \inst{4}  
        \and G. Brunetti
          \inst{2}
          \and R. Cassano
          \inst{2}
             \and F. De Gasperin
          \inst{2} 
          \and F. Gastaldello
          \inst{9}
             \and K. Knowles
          \inst{11,12}
             \and F. Loi
          \inst{10}
                   \and T. Shimwell
            \inst{5}
          \and R.~J. van~Weeren
          \inst{5}
          }

   \institute{Dipartimento di Fisica e Astronomia, Universit\`a di Bologna, via P. Gobetti 93/2 - IT 40129 Bologna, Italy 
   \and {IRA INAF, via P. Gobetti 101, IT 40129 Bologna, Italy }\\
              \email{annalisa.bonafede@unibo.it}
          \and{Université Paris-Saclay, Université Paris Cité, CEA, CNRS, AIM, 91190 Gif-sur-Yvette, France}
         \and{Hamburger Sternwarte, University of Hamburg, Gojenbergsweg 112, 21029 Hamburg, Germany }
         \and{Leiden Observatory, Leiden University, PO Box 9513, 2300 RA Leiden, The Netherlands}
         \and{Astronomisches Institut der Ruhr-Universit\"{a}t Bochum (AIRUB), Universit\"{a}tsstra{\ss}e 150, 44801 Bochum, Germany}
         \and{Ruhr Astroparticle and Plasma Physics Center (RAPP Center), 44780 Bochum, Germany}
         \and{Minnesota Institute for Astrophysics, University of Minnesota, 116 Church St. SE, Minneapolis, MN 55410, USA}
         \and{INAF - IASF Milano, via A. Corti 12, 20133 Milano, Italy}
         \and{INAF - Osservatorio Astronomico di Cagliari, via della scienza 5, Selargius, Italy}
         \and{Centre for Radio Astronomy Techniques and Technologies, Department of Physics and Electronics, Rhodes University, Makhanda, 6140 South Africa}
         \and{South African Radio Astronomy Observatory, Liesbeek House, River Park, Mowbray, Cape Town, 7700, South Africa}
             \\
             }

   \date{Received; accepted }

 \abstract{We present MeerKAT observations of the galaxy cluster Abell~2034, a massive ($M_{500}=5.21 \cdot 10^{14} \, M_{\odot}$) nearby cluster in a merging state. Previous observations at 144 MHz have shown that the cluster exhibits a plethora of diffuse emission, with multiple diffuse sources of uncertain classification because of the lack of spectral and polarimetric observations. MeerKAT multi-frequency observations, centered at 816 MHz and 1.28 GHz, together with archival low-frequency LOFAR observations at 144 MHz have allowed us to shed light on the properties of these sources. The polarization properties and spectral index information let us conclude that the cluster hosts one radio relic, a source with a very steep spectrum, previously classified as candidate relic, and filaments of very steep emission around the tailed radio galaxies identified at low frequencies.  The presence of a radio halo is confirmed, and its spectrum shows hints for curvature between 144 MHz ad 1.28 GHz.
 The polarimetric data in the L-band, together with the model of the gas density derived from X-ray observations are used to constrain the magnetic field in the intracluster medium. We assume a radially symmetric magnetic field model, whose strength declines with the cluster gas density as $B(r) \propto n_e(r)^{0.5}$, and normalize its strength within $R_{500} (B_{500})$. We find that $B_{500}=$ 1\;$\upmu$G  best explains the Faraday depth properties of the cluster, though the detection of sources close to the cluster center would be crucial to discriminate among different values.
 We conclude that the cluster Abell~2034 shows  diffuse emission with complex morphologies that do not follow the historical categories of halos and relics.
 Deep multi-frequency and polarimetric observations are fundamental to understand their origin. }


   \maketitle
%

\section{Introduction}
Galaxy clusters are massive objects that form through mergers at the intersections of the nodes of the cosmic web \citep[see e.g.][]{BorganiKravstov19}. They contain dark matter, galaxies, hot and diffuse gas that emits in the X-rays through optically thin bremsstrahlung \citep[e.g.][]{Kaastra08}, and magnetic fields that are revealed by synchrotron emission in the radio band \citep{Feretti12,vanWeeren19}. \par
Diffuse radio emission in the intra-cluster medium (ICM) is classified into several categories based on size, location, and polarization properties, which in turn are assumed to trace the different origins of these sources. \citep[e.g.][]{BJ14}. The main categories are (i) radio halos,  Mpc-size sources found in merging clusters, mainly co-spatial with the X-ray emission and likely linked to turbulent re-acceleration processes; (ii) mini-halos, found at the center of cool-core clusters, with sizes of a few hundreds kpc, possibly connected to sloshing motions of the core and/or to hadronic processes; and (iii) radio relics, Mpc-size found at the peripheries of merging clusters, polarized up to 70\%,  and linked to shock (re)-acceleration. \par 
With the advent of the LOw Frequency ARray \citep[LOFAR;][]{vanHaarlem2013}, the Murchison Widefield Array \citep[MWA;][]{Tingay2013_MWA,Wayth2018_MWA2}, the uGMRT \citep{Reddy2017_uGMRT}, and MeerKAT \citep{MeerKAT}, the view of non-thermal emission in clusters has changed:
new types of intermediate sources have  been detected, which  
make the distinction between mini-halos and halos less defined \citep{Savini_2018,Savini19,Biava21b,Biava24,vanWeeeren24}
and radio emission has been detected at the very periphery of clusters \citep{Cuciti22,Botteon22b} and even in bridges connecting galaxy clusters \citep{Govoni19,Botteon20,Pignataro24,Vernstrom21}. In addition, high-resolution radio observations show that the ICM is filled with long filamentary emission, connected to the activity of radio galaxies \citep{Brienza20,Rudnick22,DeRubeis25D}.\par
Among the clusters showing complex radio emission which does not easily fall into the classical categories, Abell~2034 is one of the first reported cases. A ``plethora of diffuse emission'' has been detected by deep observations with LOFAR \citep{Shimwell16}, but the origin of these different components was not understood because of the lack of spectral and polarization information. 

\subsection{The cluster Abell~2034, PSZ2 G053.53+59.52}

Abell~2034, also known as PSZ2~G053.53+59.52, (hereafter A2034) is a massive ($M_{500}= 5.2 \cdot 10^{14}$ \Msun), nearby ($z=0.113$) galaxy cluster \citep{Planck16}, which has been studied in the past in optical, radio, X-ray, and millimeter wavelengths. 
The main properties of the cluster are listed in Table \ref{tab:A2034}.\par
\begin{figure*}
   \centering
   \includegraphics[width=1.3\columnwidth]{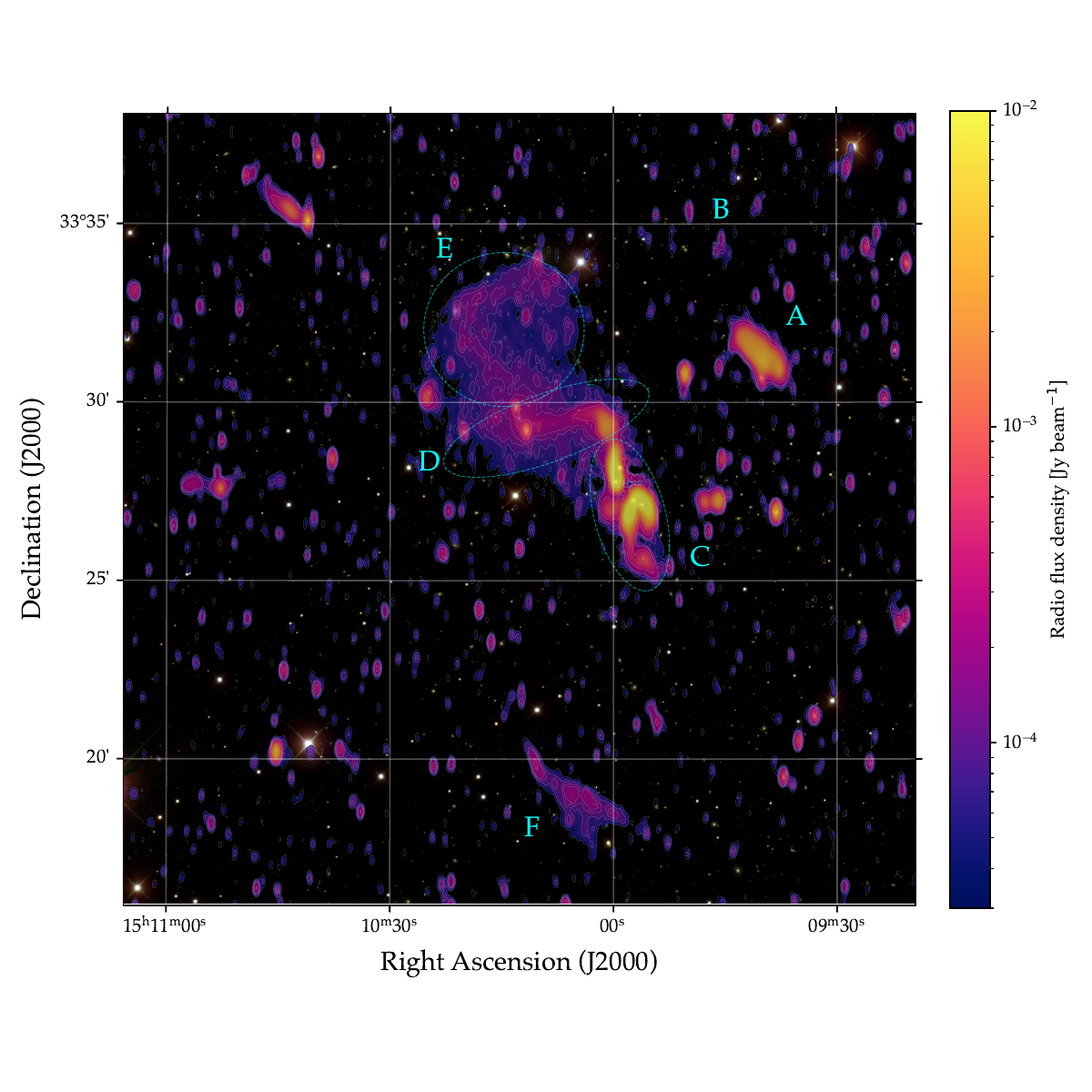}   
   \caption{The cluster Abell~2034: optical (SDSS7) and radio (MeerKAT UHF) overlay. Contours display the UHF band image, the resolution is $17.3 \arcsec \times 7.3 \arcsec$ and the noise is $\sigma_{rms}=13 \mu$Jy/beam}. Contours start at $3\sigma$ and are scaled by a factor of 2. Ellipses and text labels refer to the sources identified in \citealt{Shimwell16}.              \label{fig:A2034_X-radio}%
    \end{figure*}

A2034 is in a merging state, displaying multiple mass concentrations \citep{Owers14,Finner2025}. 
The X-ray morphology and the analysis of the morphological indicators by \cite{Campitiello22} also indicate a clear merging state. A shock with a Mach number $\mathcal M =1.59^{+0.06}_{-0.07}$ has been found at the northern edge of the cluster by \cite{Owers14}. 
\cite{Owers14} conclude that the merger axis is within 23$^{\circ}$ from the plane of the sky, along the north-south direction.
\\
A2034 is known to host radio emission, discovered by \cite{KempnerSarazin01} and studied by \cite{Giovannini09,vanWeeren11,Shimwell16}. In particular, observations at 144 MHz with LOFAR have revealed a complex mix of diffuse radio emission throughout the ICM of A2034 \citep{Shimwell16}. The main findings of \cite{Shimwell16} - having the sources labeled accordingly (see Fig. \ref{fig:A2034_X-radio}) - are:
\begin{itemize}
\item the observation of ``an irregular radio halo'' (diffuse source  E in \citealt{Shimwell16}), spatially coincident with the brightest X-ray region of the cluster, possibly characterized by a steep ($\alpha< - 1.6$) radio spectrum, estimated between 150 MHz and 1.4 GHz. Even though the authors classify the emission as a radio halo, they note that the brightness distribution of the halo is atypical and complex. 
The radio halo emission has an excess towards the north, where a shock is detected in the X-rays, and it declines towards the center. 
\item  The discovery of three candidate radio relics (source A, B, and F in \citealt{Shimwell16}), one of which, source F, is characterized by a very low surface brightness and is located at 2 Mpc from the cluster center.
\item  South of the X-ray peak (region D in \citealt{Shimwell16}), they found a bright bulb of radio emission, connected to two steep-spectrum filaments of radio emission that extend in the east-west direction, i.e. perpendicular to the merger axis. These filaments have no obvious connection to any cluster member or AGN.
\item South of the X-ray peak, connected to region D, \cite{Shimwell16} found a complex source composed of tailed radio galaxies and steep spectrum regions (complex region C in \citealt{Shimwell16}).
\end{itemize}
Overall, the nature of the non-thermal emission in A2034 remains difficult to understand without  deep multi-frequency and polarimetric observations. \par
In this paper, we present new L-band and UHF (Ultra-High-Frequency) band observations of A2034 obtained with MeerKAT, with the aim of analyzing the spectral and polarization properties of the sources detected by LOFAR and understanding their origin.\par
The paper is structured as follows: In Sec. \ref{sec:obs}, we describe the new radio observations and the data reduction process. In Sec. \ref{sec:spix} and \ref{sec:pol}, the spectral index and polarization analysis are shown.  In Sec. \ref{sec:radio}, we analyze the sources detected by LOFAR and MeerKAT observations,  and in Sec. \ref{sec:rm} we discuss the magnetic field inferred from cluster polarization properties. The discussion of the origin of the sources, following the analysis of their spectral and polarimetric properties, is presented in Sec. \ref{sec:discussion}.\par
In this paper, we adopt the convention $S(\nu) \propto \nu^{\alpha}$, with $S(\nu)$ being the flux density at the frequency $\nu$ and $\alpha$ the spectral index. We assume a $\Lambda$CDM cosmological model, with $H_0=70$ km/s/Mpc, $\Omega_M=0.3$, $\Omega_{\Lambda}=0.7$. At the cluster redshift, this gives a scale of 2.053 kpc/ \arcsec.

\begin{table}
\caption{Properties of A2034 - PSZ2~G053.53+59.52: RA, DEC, redshif, $R_{500}$ and $M_{500}$ from \cite{Planck16}.}
\begin{tabular}{ccccc}
\hline\hline 
 RA [J2000] & DEC [J200] & z & $R_{500}$ & $M_{500}$ \\
 deg       &  deg       &    &   arcmin &  \Msun \\
 \hline
227.5526   & 33.5102 & 0.113 & 9.58 &     5.21 $\cdot 10^{14}$\\
\hline\hline 
\end{tabular}
\label{tab:A2034}
\end{table}

%


\section{Observations and data reduction}
\label{sec:obs}

\subsection{MeerKAT L-band observations}
A2034 was observed in the L-band on December 24th 2021, for a total of 5h on source (DDT-20211210-AB-01).
Data have been recorded in full polarization mode, at the central frequency of 1283 MHz, with a total bandwidth of 856 MHz, divided into 4096 channels. 
The source J1939-6342 was observed  to set the absolute flux scale, following the Stevens-Reynolds 2016 flux scale \citep{StevensReynolds16_tocite}.\par
Before calibration, we have flagged the data using 
{\tt aoflagger} \citep{aoflagger}. Then, we have corrected for the handedness of the MeerKAT visibility products \citep{PerleyGreisenHugo23}
\footnote{We have used the method described here \url{https://github.com/bennahugo/ LunaticPolarimetry/blob/master/correct\_parang.py}}.
Data have been calibrated in CASA \citep{CASA}. After correcting for instrumental delays in XX and YY correlations, using J1939-6342, we have computed the phase and amplitude gains of J1939-6342. Next, we have computed the bandpass gains and the instrumental polarization leakage, again using J1939-6342. We have observed the nearby source J1609+2641 every 30 minutes as secondary calibrator. After applying the calibration tables computed above, we have solved for the amplitude and phase gains of J1609+2641, obtaining one solution for both polarizations, as the intrinsic polarization of the source is unknown and these gains will be transferred to the polarization calibrator.
Visibilities shorter than 150 $\lambda$ have been excluded, to filter out other possible extended emission in the field, and to safeguard against residual Radio Frequency Interferences (RFI), which are more prominent at short baselines. Delays, amplitude gains, bandpass, and instrumental polarization corrections derived so far have been applied to the source 3C286, which served as polarization calibrator. 
3C286 was observed in two scans of 10 min each.
We have derived the ionospheric Rotation Measure (RM) towards 3C286 for the two scans independently, using {\tt RMExtract} \citep{RMextract}. It resulted in $\sim -2.9 \pm 0.1$ rad/m$^2$ and $\sim -3.7 \pm 0.5$ rad/m$^2$ for the two scans, respectively. We have selected the scan with the lowest scatter, hence more stable ionospheric RM, for calibration.
3C286 has been self-calibrated in phase to refine the phase solutions. Then, we have computed the cross-hand instrumental delays on XY and YX, using 3C286, and finally, we have solved for the residual X -- Y phase difference, i.e. the actual polarization calibration. \par
After applying the gain tables derived so far to 3C286, we have imaged 3C286 using {\tt WSclean} \citep{wsclean}. We have produced 48 images equally spaced over the observing bandwidth in Stokes I, Q, and U, using the multi-frequency deconvolution algorithm. We have verified that the polarization intensity and the polarization angle -- after correction for the ionospheric contribution -- were consistent with the most recent model of 3C286 by \citet{HugoPerley24}.
The gain tables have been applied to the target, A2034, and the observations have been averaged by a factor of 2 in frequency. We have dropped the first and last 100 channels of the band, because of the bandpass roll-off,  and we have  compressed the data with \texttt{dysco} \citep{dysco}.
A2034 has been self-calibrated using \texttt{facetselfcal} \citep{vanWeeren2021}, which used DPPP \citep{DPPP} for calibration and {\tt WSClean} for imaging. Specifically, we performed two rounds of phase-only self-calibration using an integration time of 2 minutes and two rounds of amplitude and phase self-calibration using an integration time of 20 and 1 minute, respectively, dividing the total bandwidth into twelve segments for imaging. Solutions have been smoothed over frequency, using a Gaussian kernel of 100 MHz.\par
Directional-dependent gains have been derived towards the target and towards two off-axis sources that were limiting the dynamic range of the image directional dependent gains have been computed using an integration time of 0.5 minute in phase and 20 minutes in amplitude. We have selected a squared region of size $r \sim 2 R_{100}$, corresponding to 2$\times$19.2$^{\prime}$ and we have subtracted all the sources outside it, in order to speed up the following imaging steps (the so-called ``extraction'' procedure; \citealt{vanWeeren2021}). 
The final calibrated data set is then composed of 2048 channels.
The final image has been obtained using a Briggs weighting scheme, using robust =-0.5. It has a resolution of $6.5^{\prime \prime} \times 12.4^{\prime \prime}$ and an rms noise of 8~$\upmu$Jy/beam, which is close to the theoretical noise of 7.8$~\upmu$Jy/beam expected at this declination and with the specifics of our observations. We have re-imaged the data at different resolutions, using different weighting schemes and UV-tapers, as specified in Table \ref{tab:images}.

\subsection{MeerKAT UHF band observations}
Abell~2034 has been observed in the UHF band on December 29th 2021, for a total of 5h on source (DDT-20211210-AB-01). Observations have been taken at the central frequency of 816~MHz, and a total bandwidth of 544~MHz, divided into 4096 channels. The source J1939-6342 has been observed twice to set the absolute flux scale, and to calibrate the instrumental delays and the bandpass gains. The source J1609+2641 has been observed every 30 minutes, and used as a secondary calibrator. Data have been calibrated with the  Containerized Automated Radio Astronomy Calibration (\texttt{CARACal}) pipeline \citep{Caracal}, using a multi-component skymodel for J1939-6342. After deriving XX and YY instrumental delays, {\tt  Caracal} computes the phase and amplitude gains of J1939-6342, and corrects for the bandpass response. The first set of solutions are applied and additional flags are performed before recomputing the calibration tables. Delays and bandpass gains are applied to J1609+2641 before correcting for the amplitude and phase gains of the secondary calibrator. As done for the primary calibrator, the gain tables are applied to J1609+2641 and additional flags are done. The amplitude gains are then bootstrapped using the amplitude gains derived for J1939-6342.
The calibration tables are applied to the target Abell~2034, which has been averaged by a factor of 2 in frequency. \par
The target data have been self-calibrated using \texttt{facetselfcal} \citep{vanWeeren2021}. We have performed two rounds of phase-only (computing one solution each 5 min) and two rounds of direction-independent amplitude and phase self-calibration (computing one solution each 5 min in phase and amplitude), dividing the bandwidth into twelve channels, and smoothing the solutions in frequency with a Gaussian kernel of 100 MHz.
We performed additional rounds of directional-dependent self-calibration towards the cluster and towards off-axis sources that were responsible for artefacts on the cluster, using a solution interval of 1 min in phase and 5 min in amplitude. Sources outside a box of size equal to 2$R_{100}$ have been subtracted to speed up the following imaging steps. Data have been averaged further by a factor 4 in frequency.
The final calibrated dataset is then composed of 512 channels.
Imaging has been done with {\tt WSClean} using the multi-frequency and multi-scale deconvolution algorithm.  The final image, obtained with a Briggs weighting scheme (robust=-0.5) has a resolution of $\sim 17.3\arcsec \times 7.3\arcsec $ and a rms noise of $\sigma_{rms}=13$~$\upmu$Jy/beam, which is the theoretical noise, mostly contributed by the confusion noise expected at this frequency and resolution\footnote{see MeerKAT sensitivity calculator \tt{https://apps.sarao.ac.za/calculators/}}.
We have re-imaged the data at different resolutions, using different weighting schemes and UV-tapers, as specified in Table \ref{tab:images}.
\begin{table*}
\caption{Images of A2034 }
\centering
\begin{threeparttable}
\begin{tabular}{lccccccc}
\hline\hline 
Image    &  Freq.    &   UV range  & UVtaper & Stokes &   $\theta$ & $\sigma_{rms}$   & Figure of ref\\

           &   MHz   &               &   $\asec$  &     & $\asec \times \asec$  & $ \rm{\upmu Jy/beam}$& \\ 
           \hline
L-band   &    1284    & all & 6  &  I   & 12.4 $\times$ 6.5 & 8 &  \\
UHF band     &  816   & all &  & I & 17.3 $\times$ 7.3 & 13 &  \ref{fig:A2034_X-radio}\\
\hline
\multicolumn{8}{c}{Images for spectral index analysis}\\
\hline
L-band HR   & 1284 & $> 80  \lambda$ &  & I & 18  $\times$ 8 & 10   & \ref{fig:L-UHF_HR}  \\
UHF band HR    &  816   & $> 80 \lambda$ &     & I &   18  $\times$ 8     & 15  &  \ref{fig:L-UHF_HR} \\
L-band LR   & 1284 & $> 80 \lambda$ & 15 & I & 21  $\times$ 21 & 19  &\ref{fig:halo} \\
UHF band LR    &  816   & $> 80  \lambda$ &  15   & I &   21  $\times$ 21     &  22  & \ref{fig:Sources_check}\\
\hline
\multicolumn{8}{c}{Images for polarization analysis}\\
\hline
L-band  I  & 1284 & all  &  & I & 18  $\times$ 10 & 9 &   \\
L-band Q   & 1284 & all &  & Q & 18  $\times$ 10 & 9 &   \\
L-band U   & 1284 &  all &  & U & 18  $\times$ 10 & 9   &   \\

\hline\hline 
\end{tabular}
\begin{tablenotes}
\item Details on the images presented in this work. Col. 1: Image reference name; Col. 2: central frequency;  Col 3: UV range used in the image; Col. 4: Gaussian taper applied to the UV data; Col 5: Stokes parameter of the image; Col 6: Resolution of the restoring beam  
\end{tablenotes}
\end{threeparttable}
\label{tab:images}

\end{table*}

\begin{table*}
\caption{Flux density of Sources in A2034 }
\centering
\begin{threeparttable}
\begin{tabular}{lcccc}
\hline\hline 
Source name   &  Band.    &  S & $\alpha$ &  P \\

           &   MHz       &  mJy    & &  W/Hz \\ 
           \hline

Source A  (Relic)  & L      &   31.0   $\pm$ 1.5  & -1.3$\pm$ 0.2 &1.05$\pm 0.05  \cdot 10^{24}$  \\       
            & UHF   &  56.2 $\pm$ 5.6  & -1.3$\pm$ 0.2  &   1.9 $\pm  0.2 \cdot 10^{24}$\\
Region E (Halo)  &   L     &  23. 5$\pm$ 1.2 & -$1.8 \pm 0.3$  & 8.2 $\pm 0.2 \cdot 10^{23}$  \\ 
            &   UHF   & 61.3 $\pm$ 6.1 & -$1.8 \pm 0.3$   & 2.2 $\pm 0.2 \cdot 10^{24}$\\
Source F & L     &  1.27 $\pm$ 0.06 & -3.0$\pm$ 0.2 & 5.2 $\pm 0 .2 \cdot 10^{22}$\\
        & UHF    & 7.30 $\pm$ 0.70      &  -3.0$\pm$ 0.2 & 2.8 $\pm  0.3 \cdot 10^{23}$ \\
\hline\hline 
\end{tabular}
\begin{tablenotes}
\item  
\end{tablenotes}
\end{threeparttable}
\label{tab:sources}

\end{table*}

\subsection{X-ray observations}
\label{sec:Xray_obs}
A2034 is part of the CHEX-MATE sample \citep[][The Cluster HEritage project with \xmm\ - Mass
Assembly and Thermodynamics at the Endpoint of structure formation]{CHEX-MATE1}.
The cluster has been observed by \xmm\ using the European Photon Imaging Camera (EPIC; \citealt{turner2001} and \citealt{struder2001}) for a total of $\sim 75 $ks. We selected the observation ID 0149880101 (PI C. Sarazin) which is 26 ks long, among the three available, following the CHEX-MATE quality criteria described in Sect.~3.2.1 of \citealt{CHEX-MATE1}. These were set to ensure data homogeneity, and correspond to extraction of the radial temperature profile for each object with a $\sim 15\%$ relative error at $R_{500}$.

The procedures of data cleaning and preparation are described in detail in Sect.~3.1.1 of \citealt{bartalucci2023}, which yielded a useful exposure time of 19 ks and 11 ks for the MOS1,2 and pn cameras, respectively. From these, the azimuthal median surface brightness radial profile centered on the X-ray peak was extracted as detailed in Sect.~3.3.1 of \citealt{bartalucci2023}.

With the surface brightness profile in hand, we derived the 3D gas density profile that is used in Section \ref{sec:mag_field_sim} to model the cluster magnetic field. We used the non-parametric deprojection technique described in \citealt{croston2006,croston2008}. This technique uses the cooling factor, $\Lambda(z,T)$, to convert the emission measure radial profile to density, which depends weakly on the cluster temperature, T, and abundance z. We measured the radial dependence of these quantities using the CHEX-MATE spectroscopic analysis pipeline detailed in Sect.~3 of \citealt{rossetti2024}, and used them to derive the conversion factor.

\begin{figure*}
    \centering
    \includegraphics[width=0.45\linewidth]{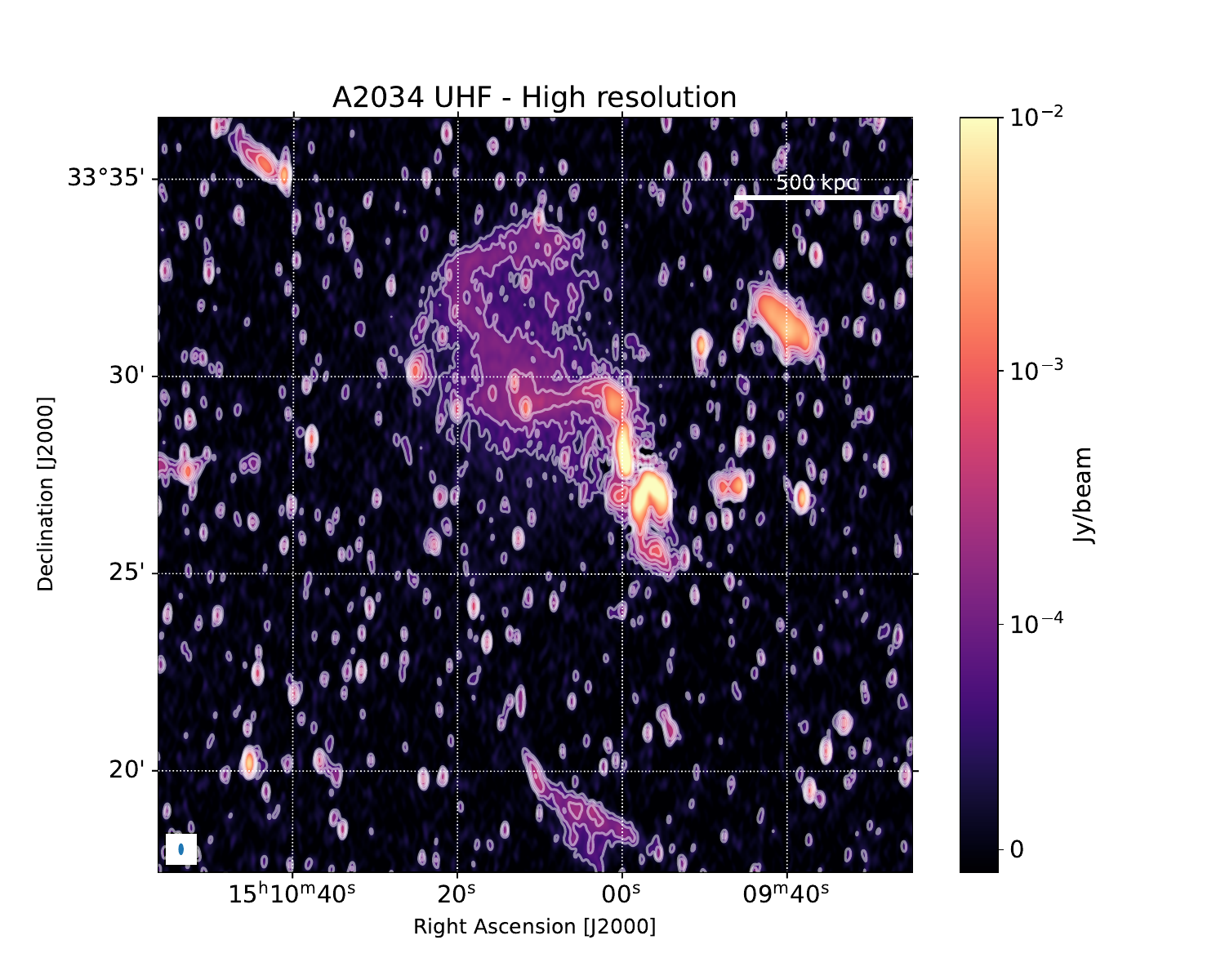}
        \includegraphics[width=0.45\linewidth]{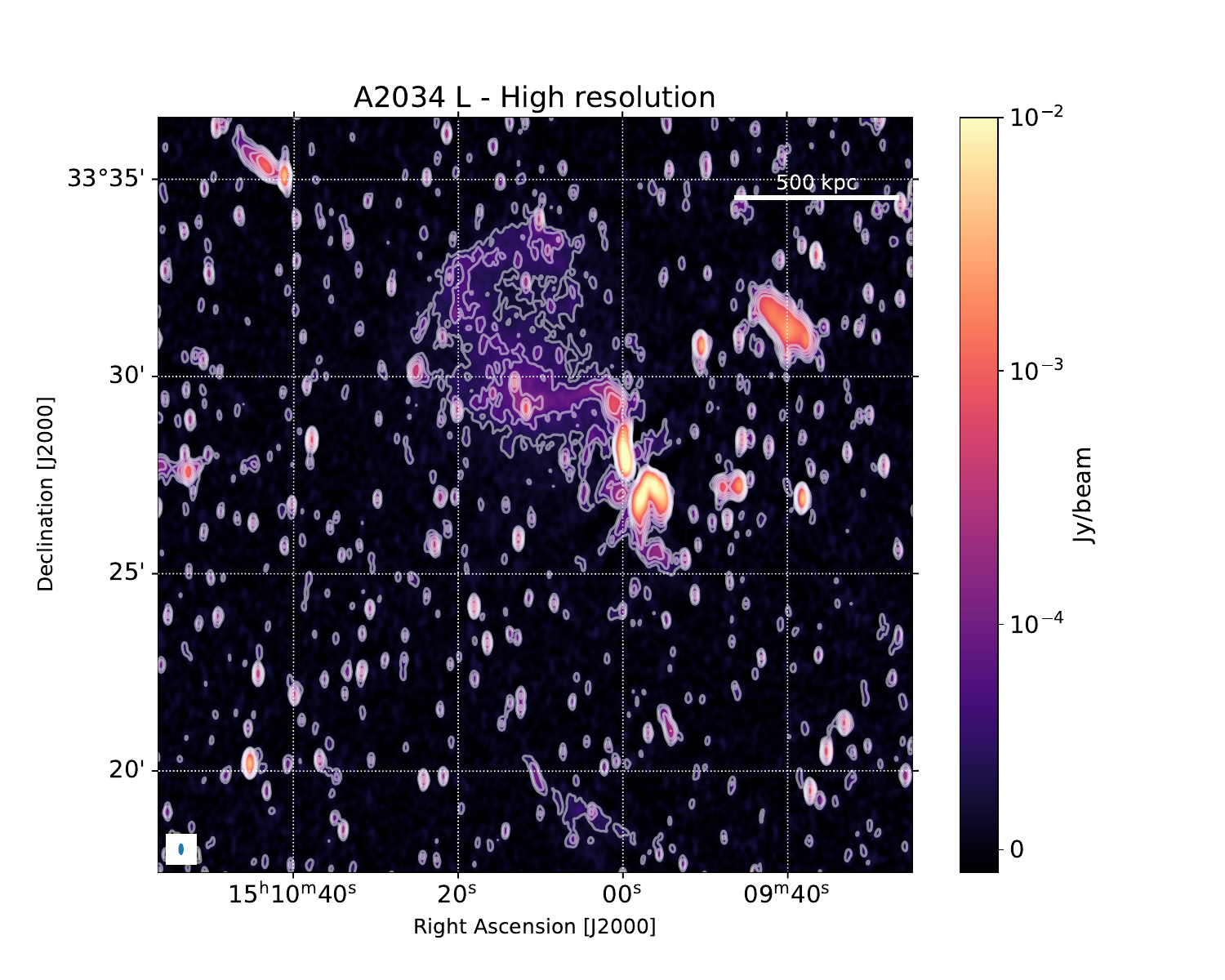}
    \caption{ The emission from A~2034 field. Left panel: UHF High resolution image in colors and contours. Contours are drawn at (3,6,12,24,48)$\times \sigma_{rms}$, with $\sigma_{rms} =15 \, \mu$Jy/beam, and the beam is 18 \arcsec $\times$ 8 \arcsec. Right panel: L band High resolution image of the same region as in the left panel. Contours are drawn at (3,6,12,24,48)$\times \sigma_{rms}$, with $\sigma_{rms} =10 \, \mu$Jy/beam, and the beam is 18 \arcsec $\times$ 8 \arcsec. }
    \label{fig:L-UHF_HR}
\end{figure*}

\subsection{Abell~2034: UHF and L-band emission}
In Fig.~\ref{fig:A2034_X-radio}, we show the continuum emission from the cluster in the UHF band, over the optical emission from SDSS7. 
In Fig. \ref{fig:L-UHF_HR}, we show the High resolution images in the L band and in the UHF band. The two images are at the same resolution, and the comparison shows how the UHF image is more sensitive to the radio diffuse emission.
We label the different components of the radio emission following \cite{Shimwell16}. 
We recover most of the features observed by LOFAR at mid- and low resolution \cite{Shimwell16}. However, no emission is detected either in the UHF band or in the L-band for source B. The emission from source D is more amorphous than in the LOFAR image, and no clear filaments are detected towards the east. Finally, only a few bright patches of the diffuse source F are detected in the L-band at high resolution, while the source is detected in the UHF band, with an extension similar to the one detected by LOFAR.
The flux density and power of the different components (A,C,D,E,F) are listed in Table~\ref{tab:sources}.\par
In Sec.~\ref{sec:radio}, we will analyze the spectral and polarization properties of the diffuse sources A-F to understand their origin.

\section{Spectral index analysis}
\label{sec:spix}
\begin{figure}
\includegraphics[width=1.1\columnwidth]{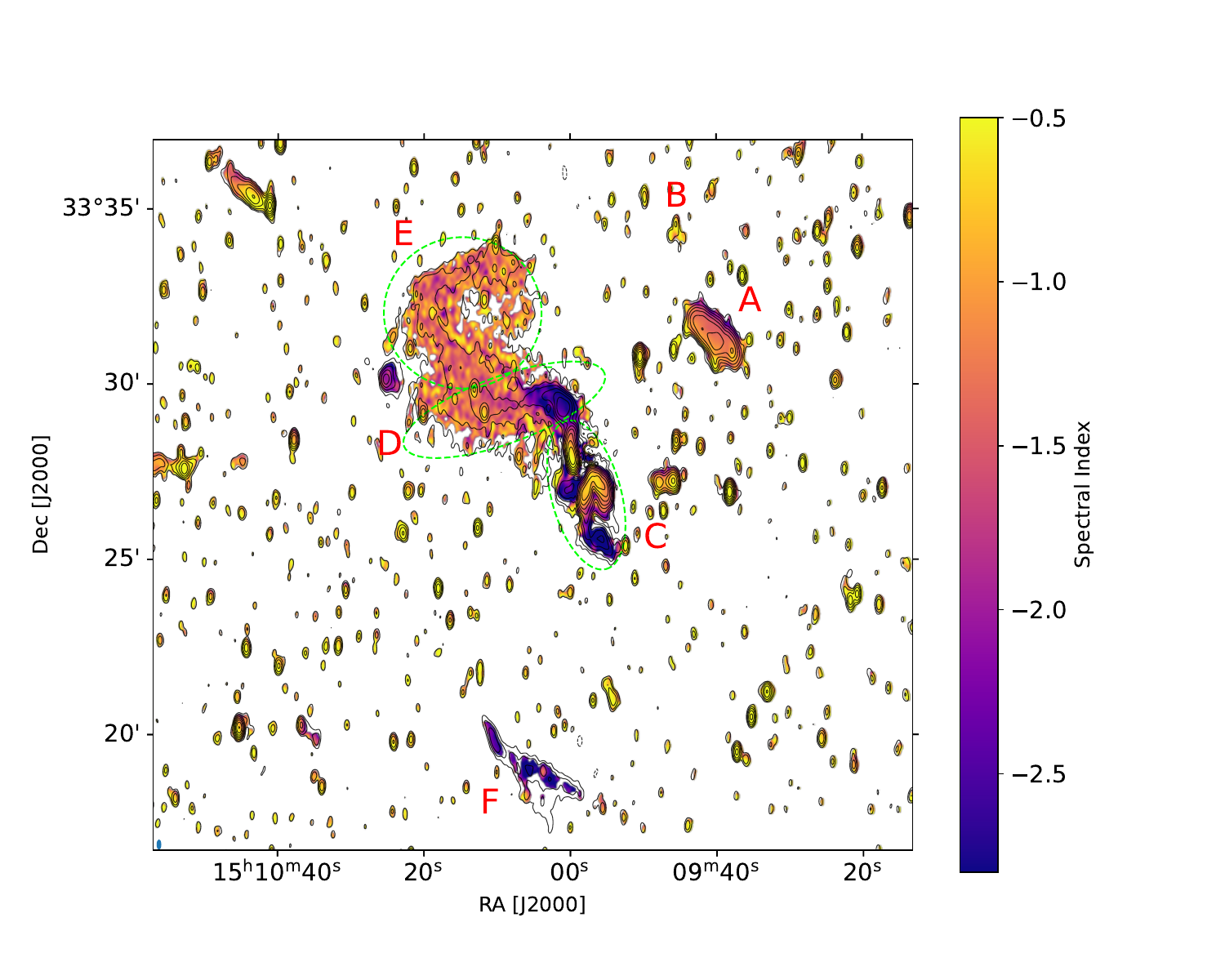}
\caption{Spectral index image of the cluster Abell~2034 computed between the L-band and UHF Band at the resolution of $18\arcsec \times 10 \arcsec$ in colors. Pixels below 3$\sigma$ in both images have been blanked. Contours refer to the UHF band at the same resolution, contours are drawn at 3,5,$\sigma$ and are then spaced by a factor 4. The rms noise $\sigma$ is 13\;$\upmu$Jy/beam. The beam HPBW is $18 \arcsec \times 8 \arcsec$. Labels refer to the sources identified in \citealt{Shimwell16}. The spectral index error in shown in Appendix \ref{appendix:spixerror}}
\label{fig:spix}    
\end{figure}

To derive the spectral information of the diffuse sources, we have combined L- and UHF band images.  Both observations have been re-imaged adopting an inner UV-cut of $80 \lambda$ to match the shortest dense sampled baselines of L-band observations. Images at both UHF and L-band have been convolved to a common restoring beam of 18\arcsec $\times$ 8\arcsec to produce the high-resolution (HR) spectral index image. We will use this image
 to derive spectral index spatial information and identify spectral index trends within sources of uncertain origin.
We have also re-imaged the data using a common UV-cut of 80$\lambda$ and a Gaussian Taper of 15\arcsec, corresponding to 13.75 k$\lambda$, in order to gain sensitivity to the low surface-brightness emission. The low-resolution images have been convolved to a common restoring beam of 21\arcsec $\times$ 21\arcsec to produce the low-resolution (LR) spectral index image. This choice is motivated by the similar resolution reached by LOFAR observations \citep{Shimwell16}.  
For both the HR and LR spectral index maps, regions below 3$\sigma$ in one of the images have been blanked in both frequency images. 
We have then computed the spectral index on a pixel basis as:

\begin{equation}
 \alpha=  \log {\frac{ S_{\rm L} / S_{\rm UHF} } {\nu_{L}/ \nu_{\rm UHF}} }
\end{equation}
and the error has been computed as

\begin{equation}
\sigma_{\alpha}= \frac{1}{\log\frac{\nu_L}{\nu_{\rm UHF}}}\sqrt{\frac{\sigma_{\rm UHF}^2}{S_{\rm UHF}^2}+\frac{\sigma_L^2}{S_{L}^2}},
\end{equation}
where $\sigma_L$ ($\sigma_{\rm UHF}$) indicate the error on the flux density $S_L$ ($S_{\rm UHF}$) in the L- (UHF) band.
The spectral index and spectral index error maps at high resolution are shown in Fig.~\ref{fig:spix} and  Fig.~\ref{fig:spix_appendix}. The LR spectral index map has been used to analyze the low surface brightness diffuse emission of source A and F (Figs.~\ref{fig:sourceA} and~\ref{fig:sourceF}).
The consistency of the absolute flux scales over the three images has been verified as we show in 
Appendix \ref{appendix:fluxscale}.

\section{L-band polarization imaging and RM-synthesis analysis}
\label{sec:pol}
We imaged the L-band data in polarization, using {\tt WSClean}. For both Stokes U and Q, we have produced 256 images over the L-band. Each image is produced averaging data from 8 channels of the UV-visibilities ( $\sim 3.3$MHz). This guarantees a sensitivity to a Faraday depth sufficiently large to map the Faraday Rotation of clusters, as better specified below (Eq. \ref{eq:phimax} \citep[e.g.][]{Govoni06, Bonafede10,Stuardi21,Vacca12,Osinga25}. \par
Stokes Q and U images have been deconvolved using the join-channel and join-polarization deconvolution algorithm of {\tt WSClean}. We also used the squared-channel-joining option, which computes the square of Q and U over the sum of the channels to find the clean components. This prevents high RM values to be averaged out when cleaning. Q and U images have been inspected, and those having a beam or rms noise larger than twice the median beam or rms noise have been discarded. The remaining Q and U images have been convolved to a common resolution of 18\arcsec $\times$ 10\arcsec and used to create Q and U cubes that are the input for the RM-synthesis analysis.\par
RM-synthesis was done using the {\tt CIRADA RM-Tools} package \citep{RMtools}.
We computed the Faraday Spectrum of the sources within a radius $r=R_{500}$ from the cluster center. The bandwidth and channel averaging performed on the data make us sensitive to a maximum Faraday depth $\phi_{\rm max}$,

\begin{equation}
|\phi_{max}| \sim \frac{\sqrt(3)}{\delta \lambda^2} = 2400~ \rm{rad/m^2},
\label{eq:phimax}
\end{equation}
while the full-width-half maximum of the Rotation Measure transfer function is:

\begin{equation}
{\delta \phi} \sim \frac{2 \sqrt(3)}{\Delta \lambda^2} = 43~ \rm{rad/m^2},
\end{equation}
giving a precision for the Faraday depth detected at 6$\sigma$ of $\sigma_{{\rm 6 SNR}} \sim ~ 3.6 \, \rm{rad/m^2}$. In the above equations, $\delta \lambda^2$ refers to the width of the channels, and  $\Delta \lambda^2$ refers to the whole observing bandwidth. The Faraday cubes have been cleaned down to 6 times the rms noise, which corresponds to a Gaussian significance level of 4.8$\sigma$ according to \citet{Hales12}. 
We have obtained the total ($F(\phi)$), real ($Q(\phi)$), and imaginary ($U(\phi)$) cleaned Faraday dispersion function (FDF) from -2000 to + 2000 rad/m$^2$, in steps of 2 rad/m$^2$. We have identified the maximum values of $F(\phi)$ for each pixel with a brightness larger than 3$\sigma_{\rm rms}$ in Stokes I and larger than 6$\sigma_{\phi}$ in $F(\phi)$.   Here, $\sigma_{\phi}$ has been computed for each pixel at the edges of the cube, in the first 180 channels of the Faraday cube, where no clean Faraday components have been found.
From here on, we refer to the maxima of $F(\phi)$  as the peaks of the Faraday spectrum.
\par
We have selected the Faraday Depth $\phi$,  corresponding to the peaks of the cleaned FDF cube, $\phi_{\rm peaks}$, to create a Faraday Depth Map, which corresponds to the Rotation Measure Map in the case of \emph{Faraday-simple} screens \citep{Brentjens05}. We will refer to this image as the RM map.\par
The ionospheric RM contribution has been computed using \texttt{RM Extract} \citep{RMextract}. Throughout the observation of the cluster, it varies from values $-2 \, \rm{rad/m^2} $ to $-4 \, \rm{rad/m^2} $. 
We have removed the average value ($-3.3 \, \rm{rad/m^2}$) and we have added to the RM error the standard deviation of the ionospheric RM ($ 0.6 \, \rm{rad/m^2}$).
The Galactic RM contribution has been removed using the Galactic Faraday depth map by \citealt{Hutschenreuter22}. Specifically, we computed the median of the values within an area of 0.25 degree squared from the center of the cluster. The median Galactic RM resulted $RM_{\rm Gal} = 3.8 \pm 3.2 ~{\rm rad/m^2}$. We note that Both the Galactic and ionospheric corrections are almost negligible with respect to the typical RM observed in clusters.\par
We have computed the polarization intensity image of the cluster, $P$ selecting the values of $F(\phi_{\rm peak})$ in each pixel of the Faraday clean spectrum, and correcting for the Ricean bias \citep{George_2012}, as $P= \sqrt{F(\phi_{\rm peak})^2- 2.3 \sigma_{\phi}}^2$.\par
The fractional polarization image of the cluster, $F_{\rm pol}$ has been obtained as $F_{\rm pol} = \frac{P}{I}$, using the Stokes I image convolved at the same resolution as the Q and U images.
 The polarization angle image has been obtained as $\psi_0 = \frac{1}{2} \arctan \frac{U(\phi_{peak})}{Q(\phi_{peak})}$ de-rotated by the cluster, Galactic, and ionospheric Faraday depths.
\par
As some regions show more than one peak in the Faraday spectrum, as we will discuss later, the values of fractional polarization should be regarded as a lower limit to the total polarization fraction.

\section{The complex radio emission from Abell~2034}
\label{sec:radio}
In this Section, we analyze the resolved spectral information and polarization properties for the two candidate relics (Source A and F) and for the 
diffuse emission at the cluster center (regions C,D, and E).

\subsection{Source A}
The diffuse source A, at the north west of the cluster, has been tentatively classified as a radio relic by \cite{vanWeeren11} and \cite{Shimwell16}.  \cite{vanWeeren11} noted that the emission could alternatively be related to the plasma from a nearby radio galaxy.
However, the lack of polarization data, the small size ($\sim$ 200 kpc), and the uniform spectral index with no clear trend towards the cluster center made the classification uncertain. \par
\begin{figure*}
    
\includegraphics[width=0.63\columnwidth]{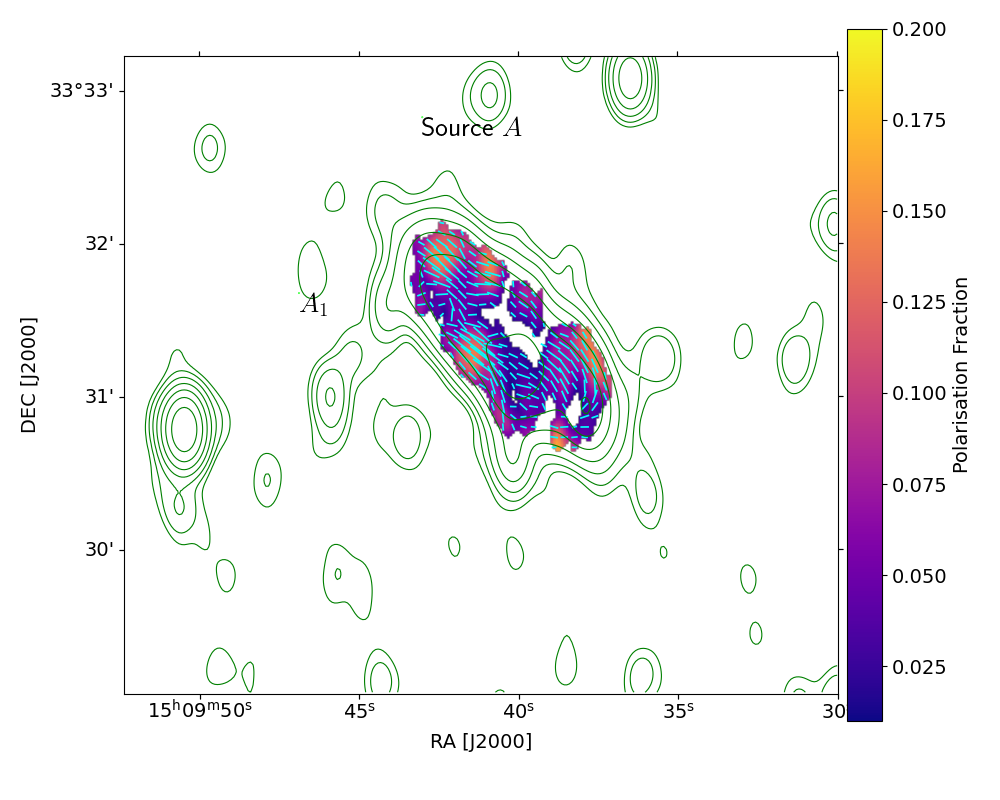}
\includegraphics[width=0.7\columnwidth]{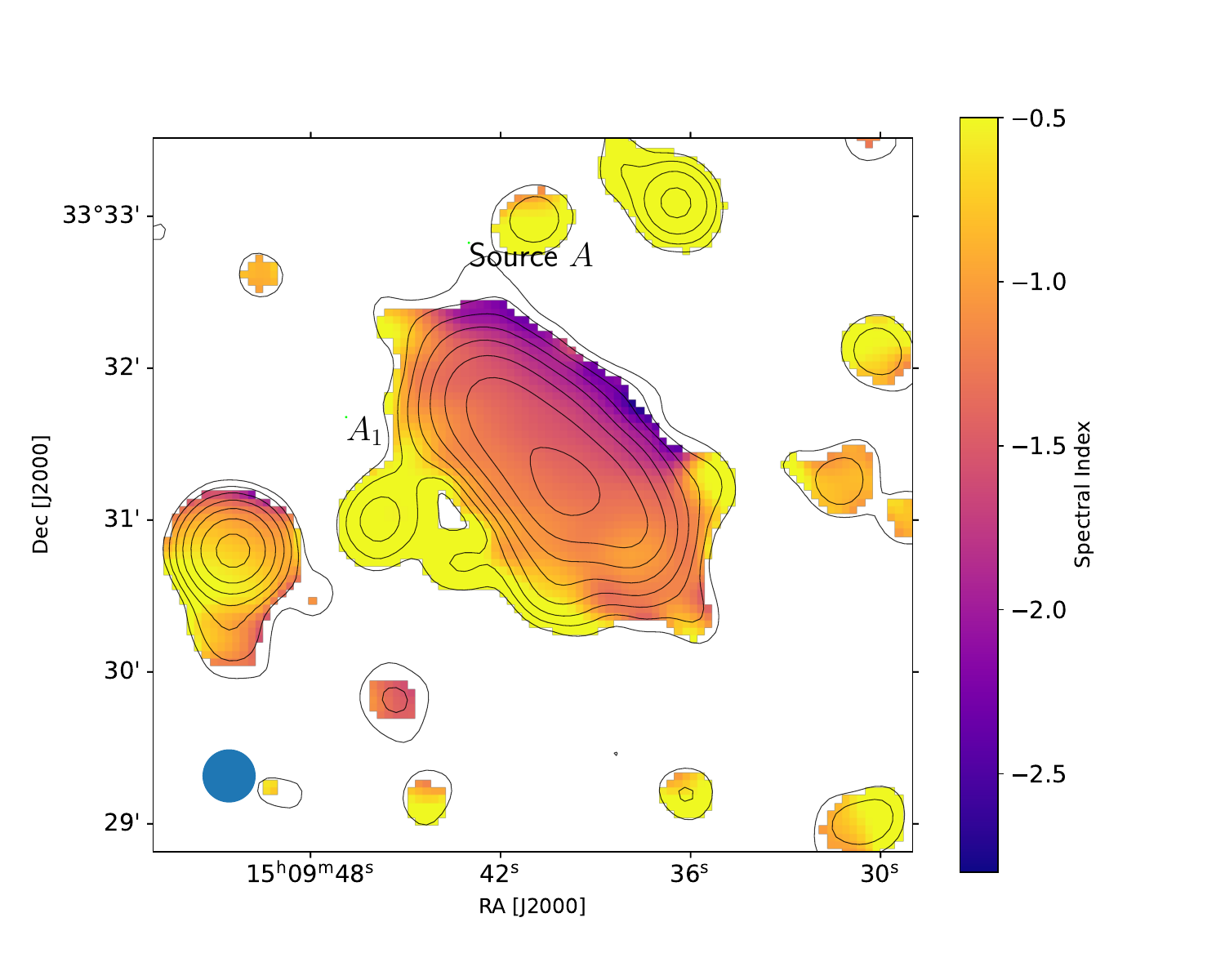}
\includegraphics[width=0.7\columnwidth]{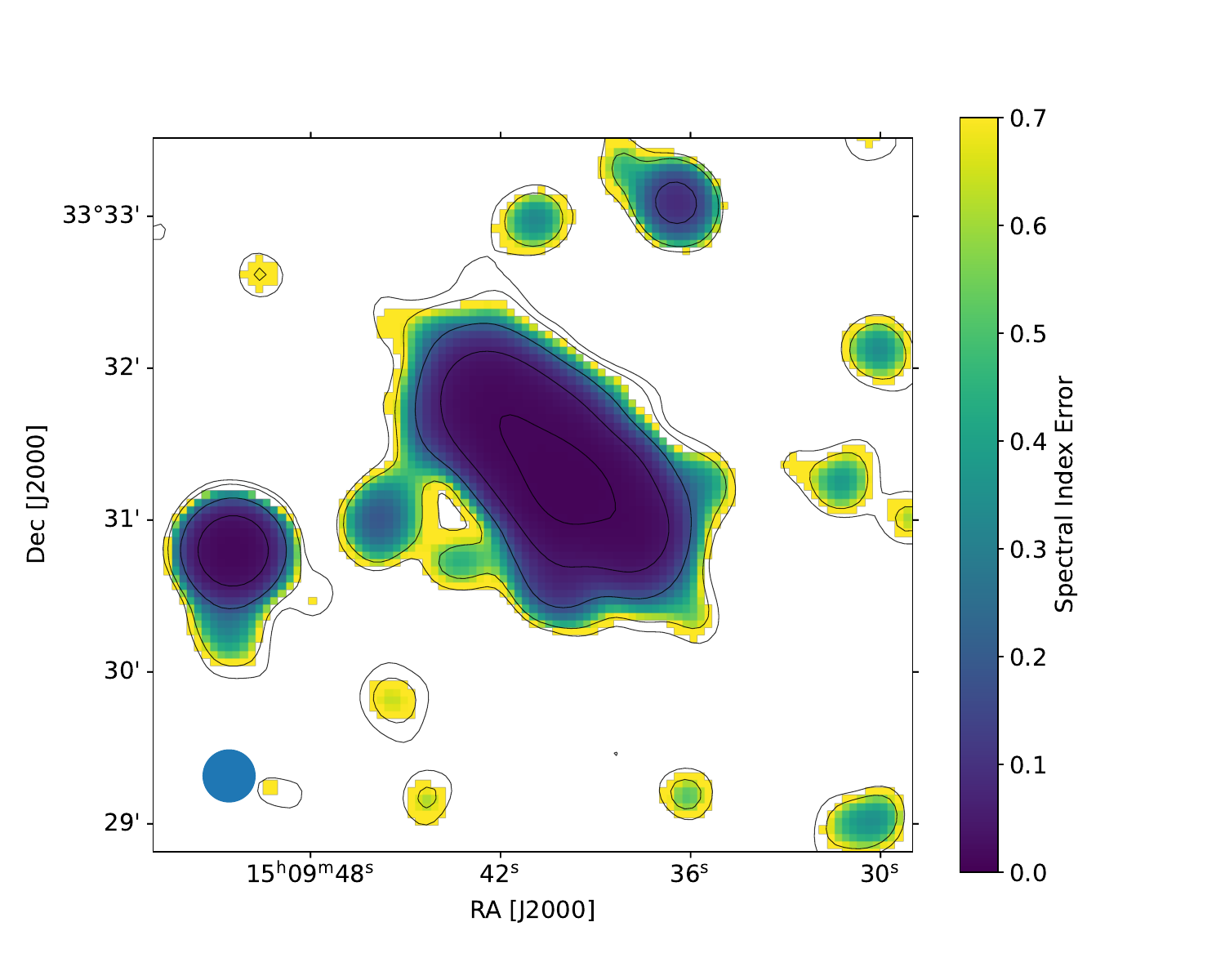}
\caption{The Source A: Left panel shows the fractional polarization in colors, the direction of the magnetic field as vectors, and the continuum emission in contours. Contours start at 3$\sigma$ and are spaced by a factor of 2. The beam is the one of the polarization images, i.e. 18\arcsec $\times$ 10 \arcsec, the rms noise is 9\;$\upmu$Jy/beam. Central panel: spectral index image computed between UHF band and L-band at Low resolution (LR images). The source $A_1$ is labeled. Contours show the emission from the UHF image. Right panel: Colors show the spectral index errors. Contours are the same as in the central panel.}
\label{fig:sourceA}
\end{figure*}

The source A is detected both in the L- and UHF band at high resolution, with an angular size of 1.9$^{\prime} \times 43$ \arcsec, corresponding to 230 kpc $\times$ 87 kpc, which is approximately the same extension as previously detected by \citet{vanWeeren11}. At low resolution, the source is more extended, measuring 2.7$^{\prime} \times 2^{\prime}$ (330 kpc $\times$ 250 kpc), and its emission is connected to the emission of a nearby radio source (source $A_1$ in Fig.~\ref{fig:sourceA}), located at the east of source A, towards the cluster center. This emission is not detected in LOFAR, WSRT, and VLA images \citep{Shimwell16,vanWeeren11,Giovannini09}. The average spectral index computed between UHF and L- band at low resolution, in the region above 3$\sigma$ in both images is $\alpha=-1.3 \pm 0.2$.\par
The spectral index trend shows a steepening towards the cluster outskirts, going from $\alpha=-0.6\pm 0.4$ in the south-east, i.e. towards the cluster center, to  $\alpha=-2.2 \pm 0.2$ in the north-west, i.e. farther out from the cluster center.
A hint for such steepening was pointed out by \cite{Shimwell16}, but the presence of a contaminating radio source and the coarser resolution did not allow the authors to draw a firm conclusion. The emission of the source $A_1$ has a spectral index $\alpha = -0.6 \pm 0.3$ and is connected to the source A through a patch of emission with a similar spectral index, despite the higher uncertainties due to the low brightness, i.e.  $\alpha= -0.6 \pm 0.5$. \par
Hence, the spectral index shows a gradient along the major axis, as expected in radio relics. But, contrary to expectations, the spectral index steepens towards the cluster outskirts. \par 
The hypothesis that source A is a radio relic can be further tested by looking at its polarization properties. Relics are known to be polarized at the 10-70 \% level in the L-band, and the magnetic field is expected to be aligned with the shock front, along the relic's main axis. In Fig.~ \ref{fig:sourceA}, the polarization fraction and polarization angle images are shown. Source A has a mean fractional polarization of $F_P \sim 7 \%$, reaching a maximum of 13\% and a minimum of 2\%. 
The polarization vectors indicate a good alignment of the magnetic field with the main axis of source A. This feature is consistent with the hypothesis that source A is a radio relic, where the shock responsible for particle (re)-acceleration has compressed the magnetic field along the shock front. The FDF of source A shows a single peak, indicating that the emission is rotated by an external Faraday screen. 
The mean RM value over the source, after correction for the Galactic foreground, is $\langle RM \rangle =- 0.8 \pm 2.0  \, {\rm rad/m^2}$, while the standard deviation $\sigma_{\rm RM}$ is 
$\sigma_{\rm RM}= 11 \pm 1 \, {\rm rad/m^2} $. Such low values of $\langle RM \rangle $ and $\sigma_{\rm RM}$ are expected for sources located at the cluster periphery \citep[e.g.][]{Bonafede13,Stuardi22}.
\par
Our spectral index and polarization analysis strengthens the hypothesis that source A is a radio relic, though the projected spectral index trend appears inverted with respect to other radio relics. Projection effects could play a role, as the shock could be moving towards the observer, at a small angle with respect to the line of sight. Hence, we could be observing the post-shock region projected towards the cluster outskirts. The connection with source $A_1$ suggests that seed electrons for (re)-acceleration could be supplied by this source. 
The magnetic field orientation, as derived from polarization, indicates that the plasma has been compressed and the magnetic field has been ordered along the presumed shock front. The fractional polarization is lower than what has been observed in other radio relics, in agreement with the idea that the shock is moving at a small angle with respect to the line of sight, causing depolarization. The Faraday spectra of the relic are consistent with one single Faraday component, however higher resolution in Faraday space, achievable with lower frequency observations, could show more complex Faraday spectra and indicate an internal depolarization of the emission..\par
If source A is powered by DSA, we can compute the shock Mach number from the radio spectral index. Under quasi-stationary conditions, we can write
\begin{equation}
M = \left(  \frac{2 \alpha_{\rm inj}-3}{2  \alpha_{\rm inj +1}}\right)^{0.5} ,
\end{equation}
where $\alpha_{\rm inj}$ is the injection index, which is linked to the integrated spectral index  $\alpha_{\rm int}$ as  $\alpha_{\rm inj}=\alpha_{\rm int}+0.5$ (see e.g. \citealt{BlandfordEichler87}). 
In the case of source A, we derive $M = 1.59 \pm 0.07$, consistent with what is expected for merging shocks \citep[e.g.][]{Vazza11,Kang12,Wittor21}.

\subsection {Source F}
Source F is detected by \cite{Shimwell16} with LOFAR at 144 MHz. It is a faint radio source, having a major axis of 290\arcsec (600 kpc) and a minor axis of 80\arcsec(165 kpc). The WSRT image could barely detect the source F, hence a limit to its spectrum was derived by measuring the flux density in the area corresponding to the LOFAR detection: $\alpha < -2.15$. \citet{Shimwell16} concluded that this source could be a radio relic, though polarization information and resolved spectral index images are needed to classify the source.\par
\begin{figure*}
    
\includegraphics[width=0.64\columnwidth]{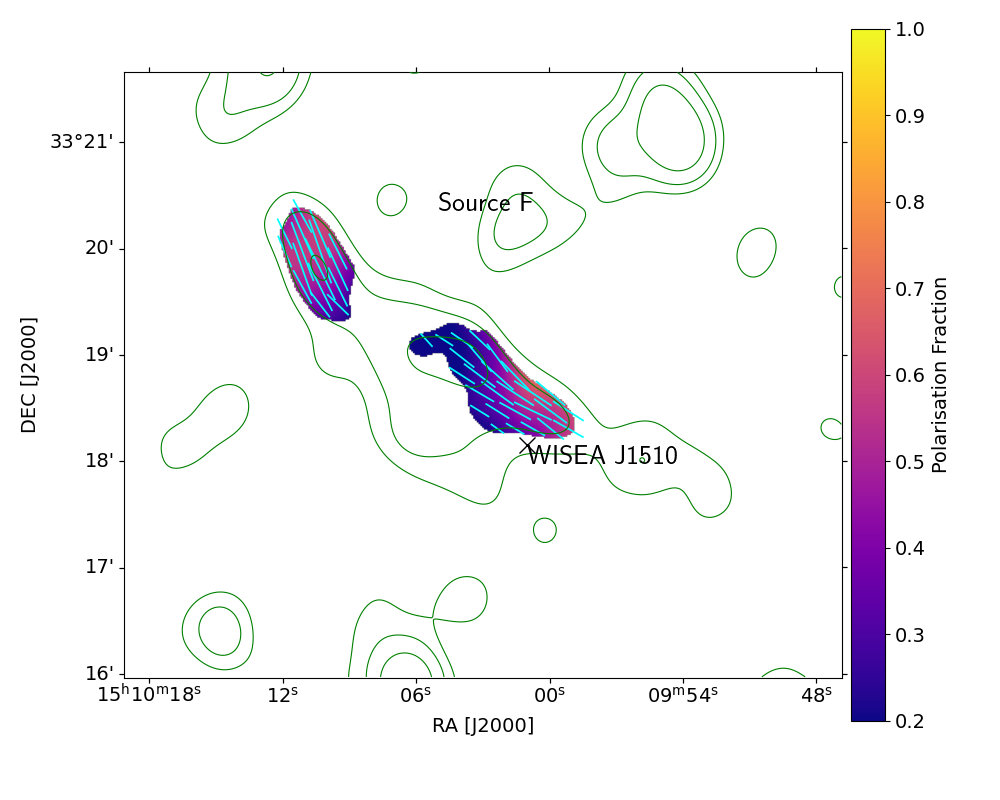}
\includegraphics[width=0.7\columnwidth]{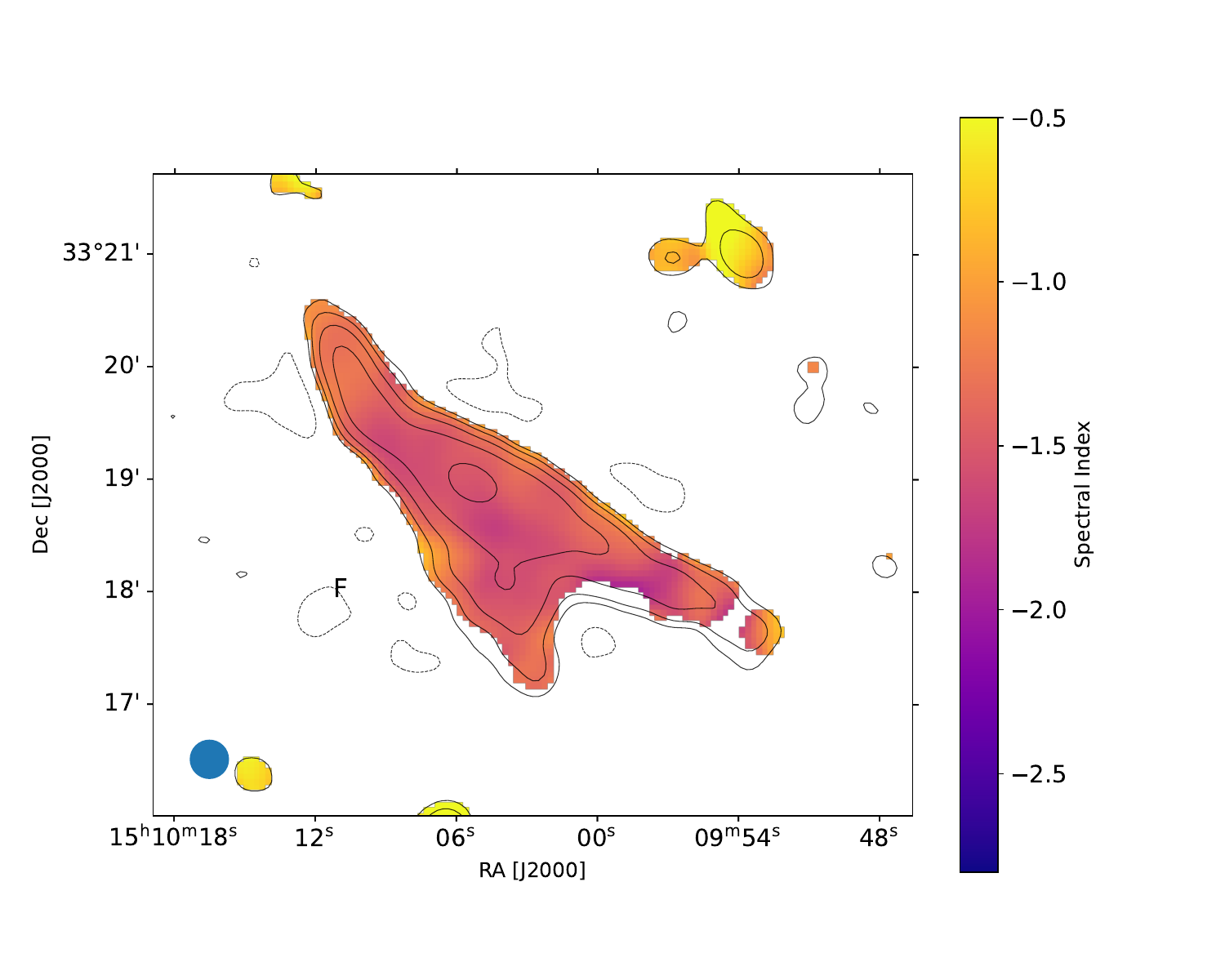}
\includegraphics[width=0.7\columnwidth]{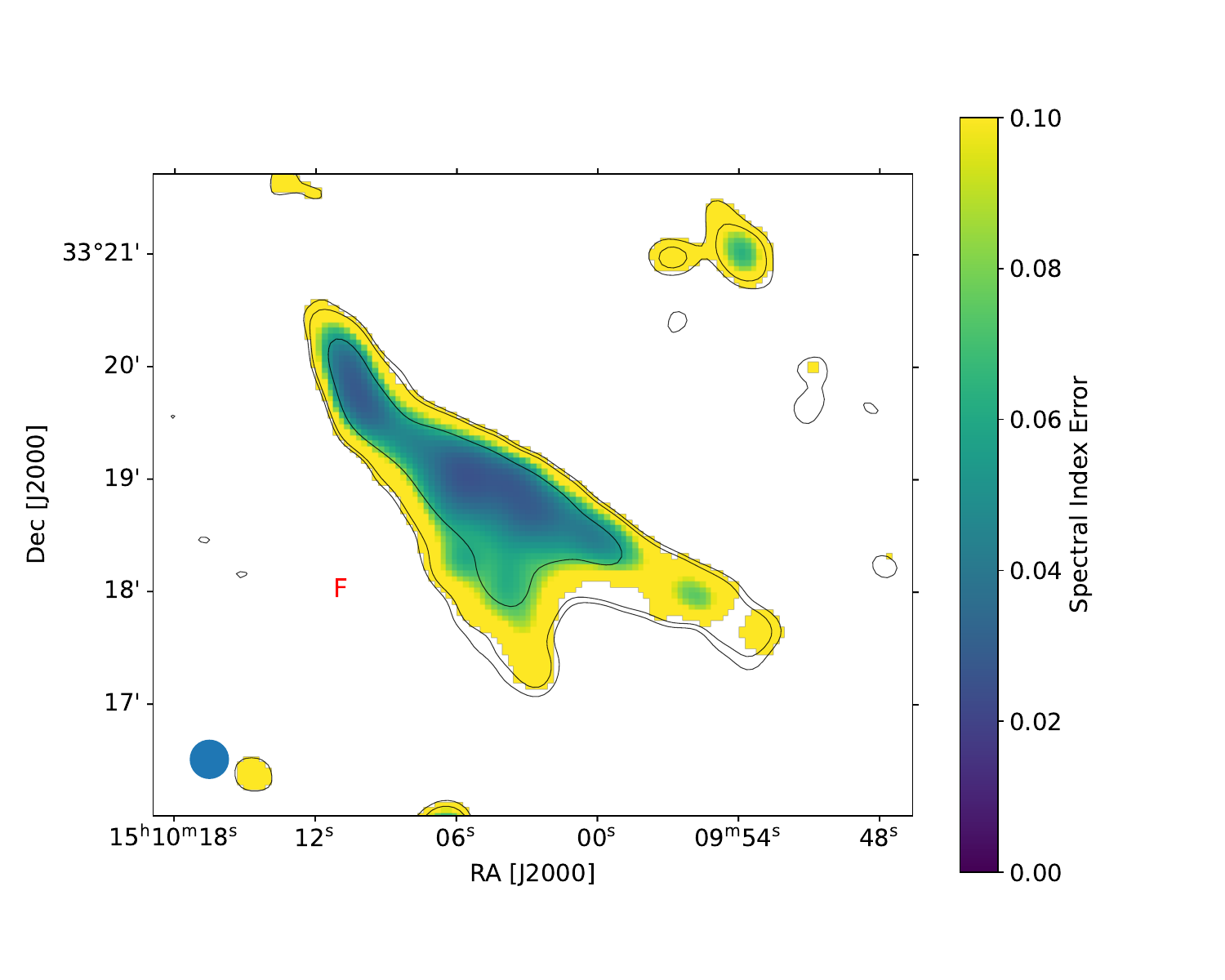}
\caption{The Source F: Left panel shows the fractional polarization in colors, the direction of the magnetic field as vectors, and the continuum emission in contours, taken from L-band images at low resolution. Contours start at 3$\sigma$ and are spaced by a factor of 2. The beam is the one of the polarization images, i.e. 21\arcsec $\times$ 21 \arcsec, the rms noise is  30$\upmu$Jy/beam. Central panel: spectral index image computed between LOFAR band and UHF band at Low resolution (LR images). Contours show the emission from the UHF image. Right panel: Colors show the spectral index errors. Contours are the same as in the central panel.}
\label{fig:sourceF}
\end{figure*}

Source F is detected in the UHF band images, located at  11.7$^{\prime}$ (1380 kpc) from the cluster center, and having a size slightly larger than observed at LOFAR frequencies (major axis 324\arcsec - 660 kpc, minor axis 95\arcsec - 200 kpc), though two point-like sources are visible in UHF, at the south-west end of the relic, embedded in the diffuse emission at low resolution.
In the L-band, the source is partially detected at low resolution. 
In Fig.~\ref{fig:sourceF}, we show the spectral index image between LOFAR and UHF band, which does not reveal any particular trend along the source main axis, as we would expect for radio relics.\par
We have estimated the spectral index using LOFAR and UHF images at low resolution. It results in $\alpha=-1.5 \pm 0.2$. Over the same area, considering a 1-$\sigma$ upper limit for the L-band flux, we obtain $\alpha < -2.2$, which is consistent with the limit reported by \cite{Shimwell16}. These values indicate a steepening of the radio spectrum between UHF and L-band frequencies.\par
To better analyze the spectral properties of the source, we show in Fig. \ref{fig:halo}, right panel, the spectral index computed between UHF and L band versus the spectral index computed between LOFAR and UHF (the so-called ``color-color plot''; \citealt{Color-color_plot}). Each point in the plot refers to the spectral index computed in boxes having an area of seven beams, and having a flux density larger than $3\sigma$ in all the 3 images. The source F lies below the power-law line (where $\alpha_{UHF}^{L-band} = \alpha _{LOFAR}^{UHF}$). For reference, we overplot the track in the color-color plot expected for an electron population aging according to the Jaffe-Perola model \citep[JP][]{JP}, assuming an injection $\alpha=-0.5$. Source F lies on the JP model line, suggesting that we are looking at aged emission from a radiogalaxy. A possible optical counterpart could be the source WISEA~J151001.00+331809.0, at $z=0.112$ \citep{WISE} \par
 We have used low-resolution polarization images to study the polarization properties of the source. The process described in Sec. \ref{sec:pol} has been applied to LR images in the L-band. After removal of the Galactic and ionospheric contribution, the Faraday depth of the source is $\langle RM \rangle = -4 \pm 1 \, {\rm rad/m^2}$, and its dispersion is $\sigma_{\rm RM}=3.7 \pm 0.5\, {\rm rad/m^2} $, as expected for sources located on the cluster outskirts.
The magnetic field, as indicated by the polarization vectors, appears ordered along the source's main axis, as expected for radio relics, though only a small fraction of the source is detected in polarization in the L-band.\par
These results on LR polarization images are likely to be affected by the Faraday emission from the Galaxy, which is detected in the LR images, as explained in Appendix \ref{appendix:GalacticRM}. We cannot unambiguously separate the polarized emission from the source from the polarized emission of the Galaxy. Hence, though the source F appears highly polarized, we cannot determine whether the detected polarization is entirely associated with the source.
\par
Our spectral index and polarization analysis reveals peculiar features of source F. Though its location and morphology resemble a radio relic, the steep spectrum and the absence of a spectral index gradient across the source main axis cannot easily be explained by shock (re)-acceleration. Its properties point towards aged plasma, possibly compressed by a shock wave.

\subsection{Source B}
No emission corresponding to source B in \citet{Shimwell16} is detected either in UHF or in L-band images. We can estimate a limit on the average spectrum, based on the average surface brightness detected in the LOFAR HBA image above 5$\sigma$, and on the 3$\sigma$ noise level of the L-band image at the same resolution. We derive $\alpha < -1.4 $. Hence, the non-detection of source B could be due to its low surface brightness and not to its very steep spectrum. We note that the upper limit on the spectrum we have derived points towards aged plasma rather than emission from a relic.

\subsection {The complex region C}
\begin{figure*}
    \centering
\includegraphics[width=0.64\columnwidth]{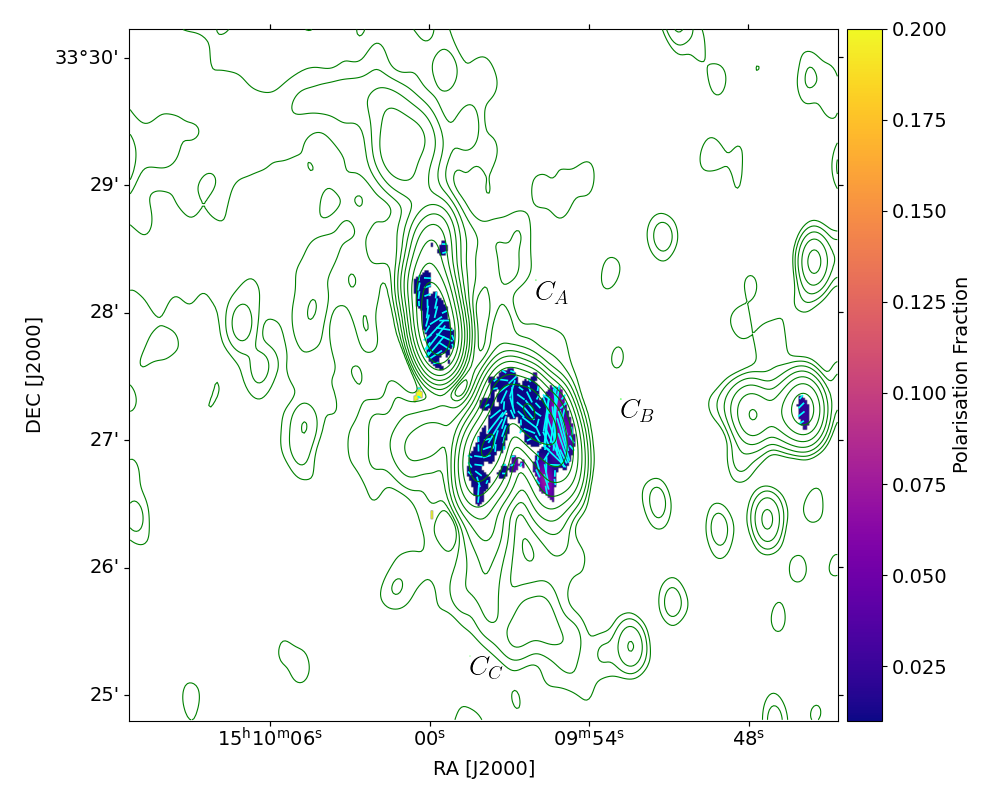}
\includegraphics[width=0.65\columnwidth]{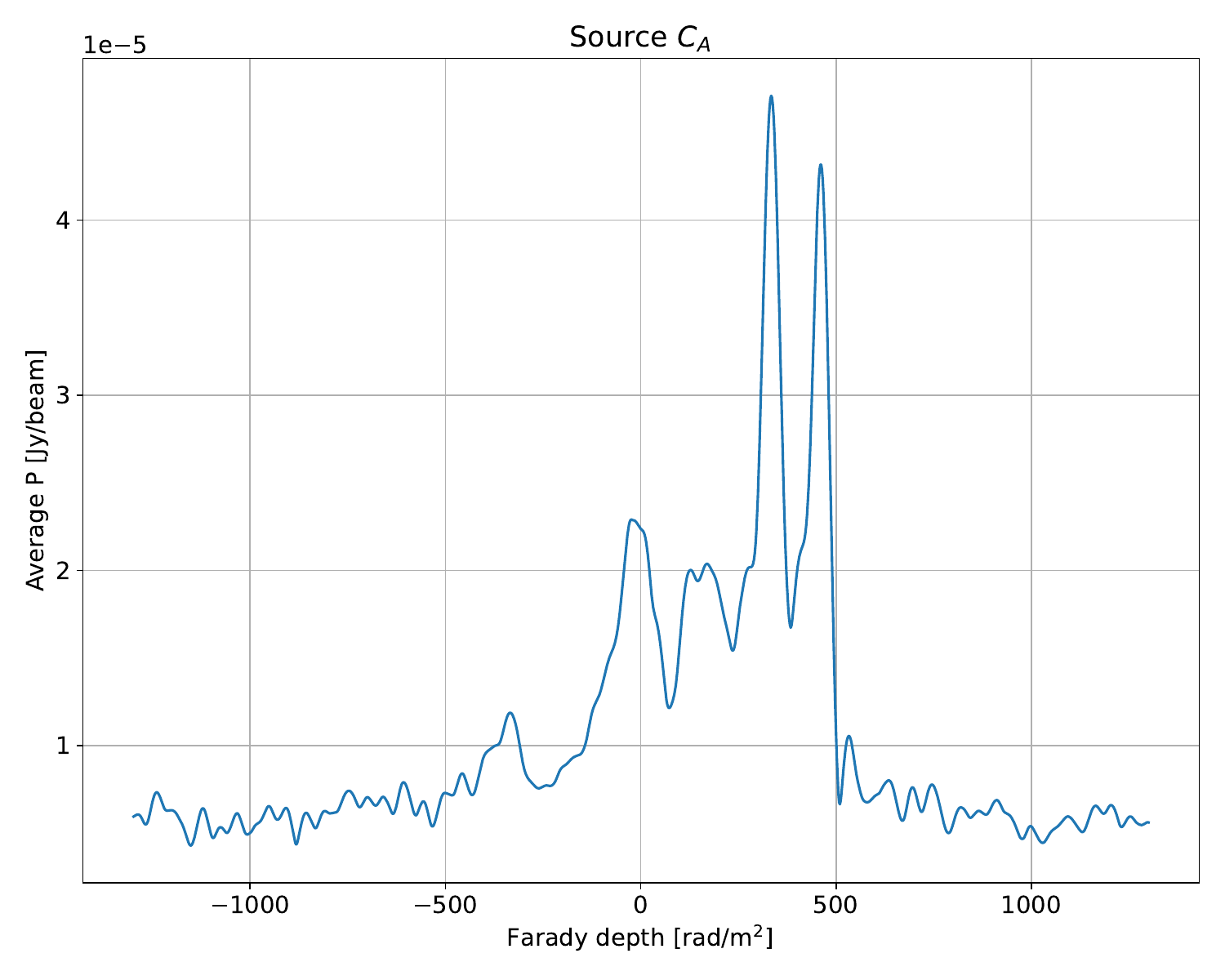}
\includegraphics[width=0.7\columnwidth]{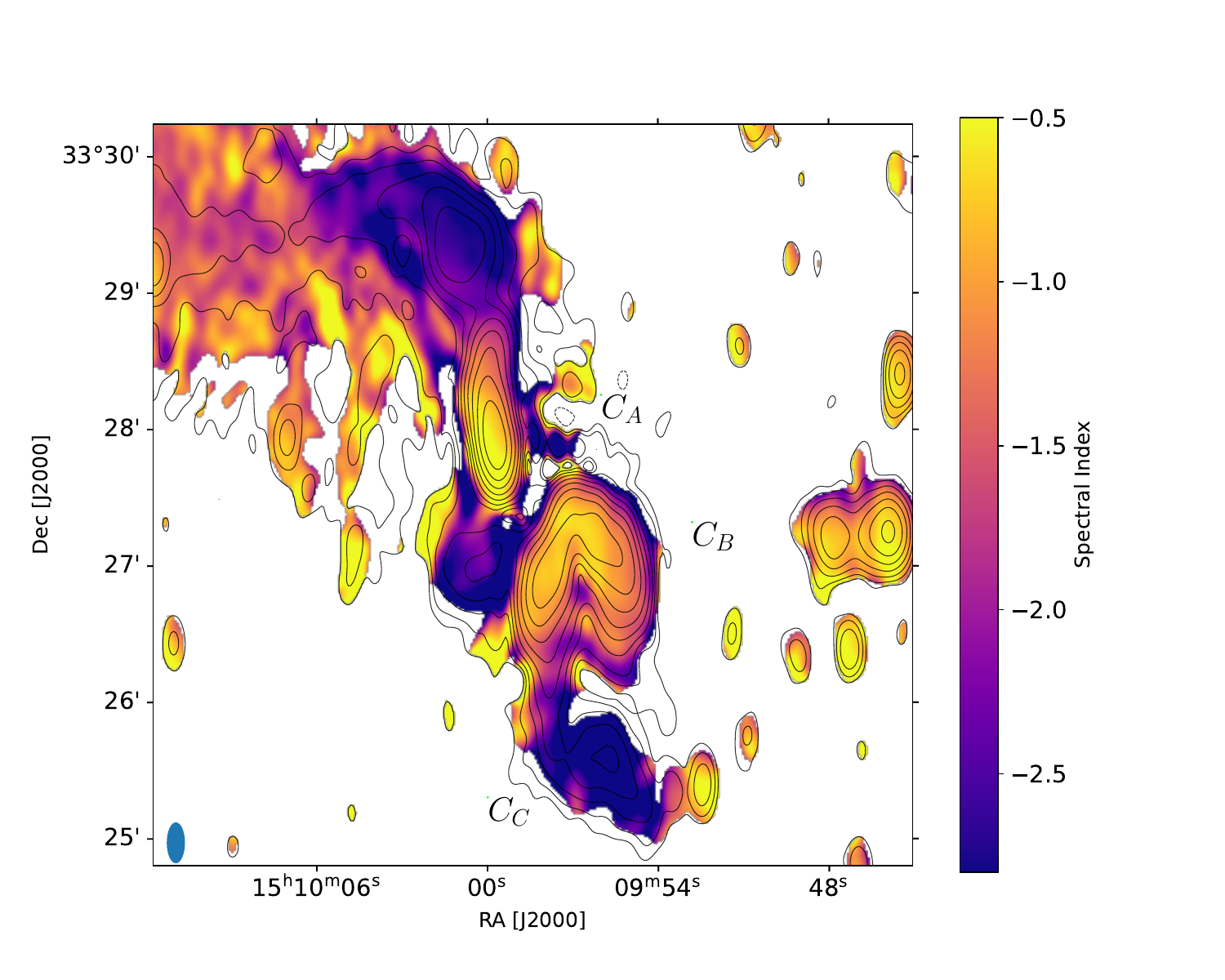}
\caption{The complex region C. Left panel: polarization fraction (in colors) as derived by the peak of the Faraday spectrum. Lines indicate the direction of the magnetic field as vectors, after rotation by the observed Faraday depth. The continuum emission in L-band in shown in contours. Contours start at 3$\sigma$ and are spaced by a factor of 2. The beam is  18\arcsec $\times$ 10 \arcsec, the rms noise is 9\;$\upmu$Jy/beam. Middle panel: averaged Faraday Spectrum of source $C_A$. Right panel: spectral index image computed between UHF band and L-band at high resolution (HR images). Contours show the emission from the UHF image.}
\label{fig:sourceC}
\end{figure*}

The complex emission south of the cluster (region C; \citealt{Shimwell16}) is detected in both the L-band and the UHF band images. Two tailed radio  galaxies with a cluster optical counterpart have been detected in LOFAR images. These sources are labeled as $C_A$ and $C_B$ in Fig.~\ref{fig:sourceC}, following \citet{Shimwell16}. Around these galaxies, very steep spectrum emission has been detected. In particular, the emission labeled $C_C$ connects to the eastern tail of $C_B$, extending southwards.\par
The radio galaxies $C_A, C_B$, and the steep-spectrum emission labeled $C_C$ are all detected in the L-band and UHF images. In Fig.~\ref{fig:spix}, a spectral index map of the cluster at high resolution is shown and in Fig.~\ref{fig:sourceC} a zoom into the complex emission region C is shown.
The tailed radio galaxy $C_A$ shows a spectrum with $\alpha =  -0.65 $  at its center. This spectrum gradually steepens along the tail, towards the North, reaching values down to $\alpha \sim -1.80 \pm 0.01 $ at the tip of the tail. Further north, the spectral index continues to steepen, as the emission from the tails blends into another blob of emission located at the north of $C_A$ and labeled $D_A$ (see Fig. \ref{fig:halo}), where the spectral index reaches values as steep as $\alpha \sim -2.8 \pm 0.1 $.\par
The radio galaxy $C_B$ shows the typical spectral index trend of tailed radio galaxies. We measure $\alpha \sim -0.6$ at the position which likely corresponds to the core. The spectral index then steepens along the two tails, reaching values $\alpha \sim -1.41 \pm 0.03$ at the end of the western tail. The eastern tail extends further south and the spectral index steepens further, reaching a very steep spectrum $\alpha \sim -3.1 \pm 0.1$ in source $C_C$. A region of steep spectrum is detected to the east of $C_B$ and to the south of $C_B$. Here, the spectral index reaches values $\alpha \sim -3.9 \pm 0.3$. 
In Fig. \ref{fig:halo}, we show the resolved spectral properties of the emission in region C in the color-color plot. The spectral index in each region is computed as explained for source F above. Some regions follow the JP track, while the steepest regions of the source C are likely aged radio plasma, observed at frequencies higher than the synchrotron cutoff frequency.
\par
The radio tails $C_A$ and $C_B$ are weakly polarized, with an average fractional polarization of $F_P \sim 2\%$. Such a low polarization fraction is expected for sources located within galaxy clusters, as their emission is depolarized by the turbulent magnetized ICM. The sources $C_A$ and $C_B$ have a projected distance of 230\arcsec (472 kpc) and 280 \arcsec (575 kpc), respectively, with respect to the cluster center. 
Indeed, the values of  $\sigma_{\rm RM}$ are higher than measured for the relic source A.
We derive for source $C_A$ $\langle RM \rangle = 260 \pm  48   \, {\rm rad/m^2}$  $\sigma_{\rm RM} = 167 \pm 24  \, {\rm rad/m^2}$
and for Source $C_B$ $\langle RM \rangle = =- 11 \pm  12   \, {\rm rad/m^2}$, $\sigma_{\rm RM} = 43 \pm 6  \, {\rm rad/m^2}$. 
The source $C_A$ displays at least two separate peaks in the FDF, indicating multiple emission regions along the line of sight. In Fig.~\ref{fig:sourceC}, the average FDF of the source is shown. 
A more detailed description of the RM  and its radial profile will be presented in Sec. \ref{sec:rm}.

\subsection {Diffuse sources D and E}
The diffuse emission at the center of the cluster has been classified as a radio halo by \citet{Shimwell16}. 
The authors distinguish the emission at the center of the cluster in diffuse region E, which is the diffuse and low-brightness emission co-spatial with the brightest X-ray emission, and region D, which corresponds to a bright filament of emission, south of region E (see Fig.~\ref{fig:A2034_X-radio}). This bright filament of emission, detected with LOFAR, departs from a bright bulb of emission, located at 15:10:00 +33:29:20 (source $D_A$), and extending for $\sim$ 0.5 Mpc. \par
\begin{figure*}
\includegraphics[width=\columnwidth]{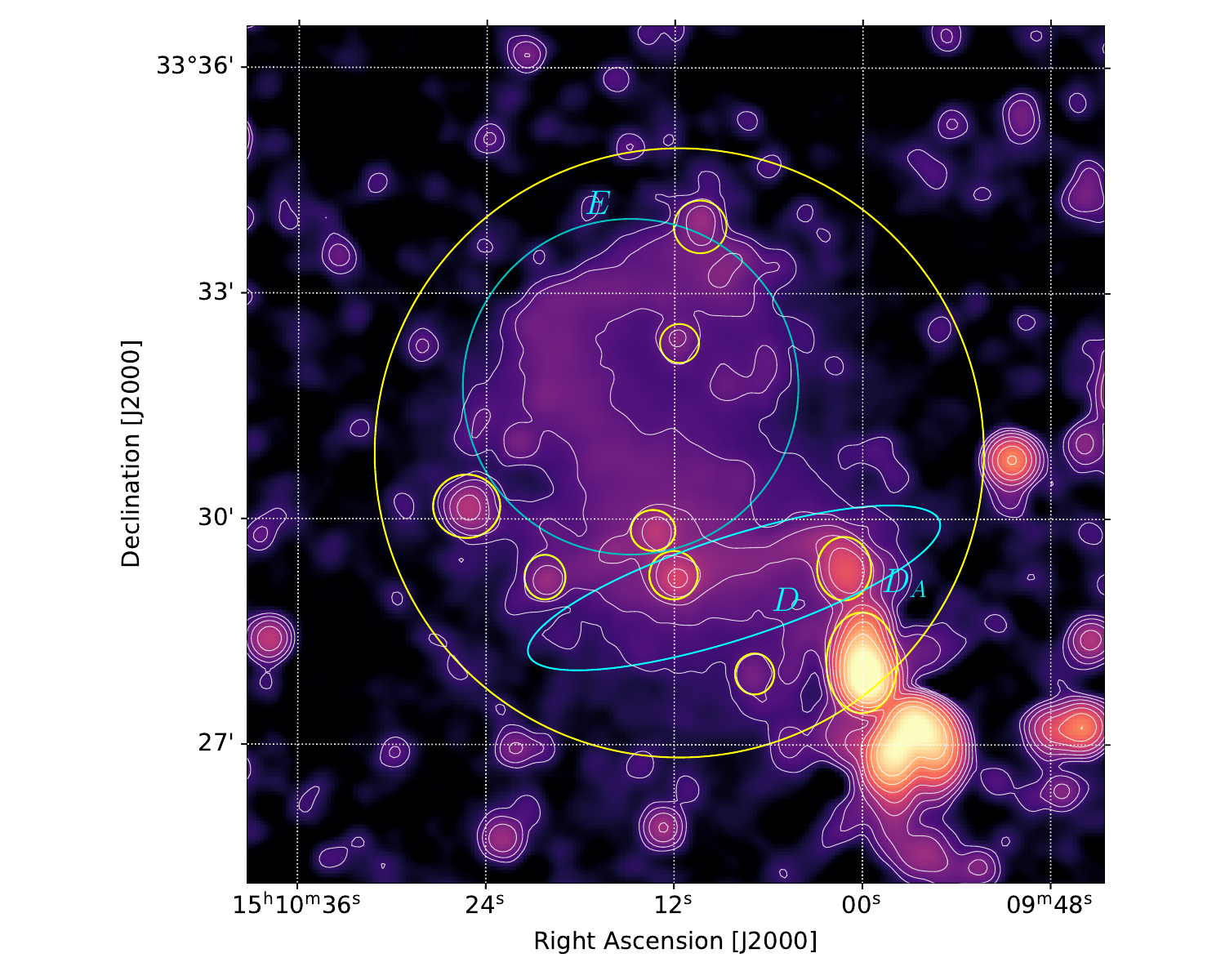}
\includegraphics[width=\columnwidth]{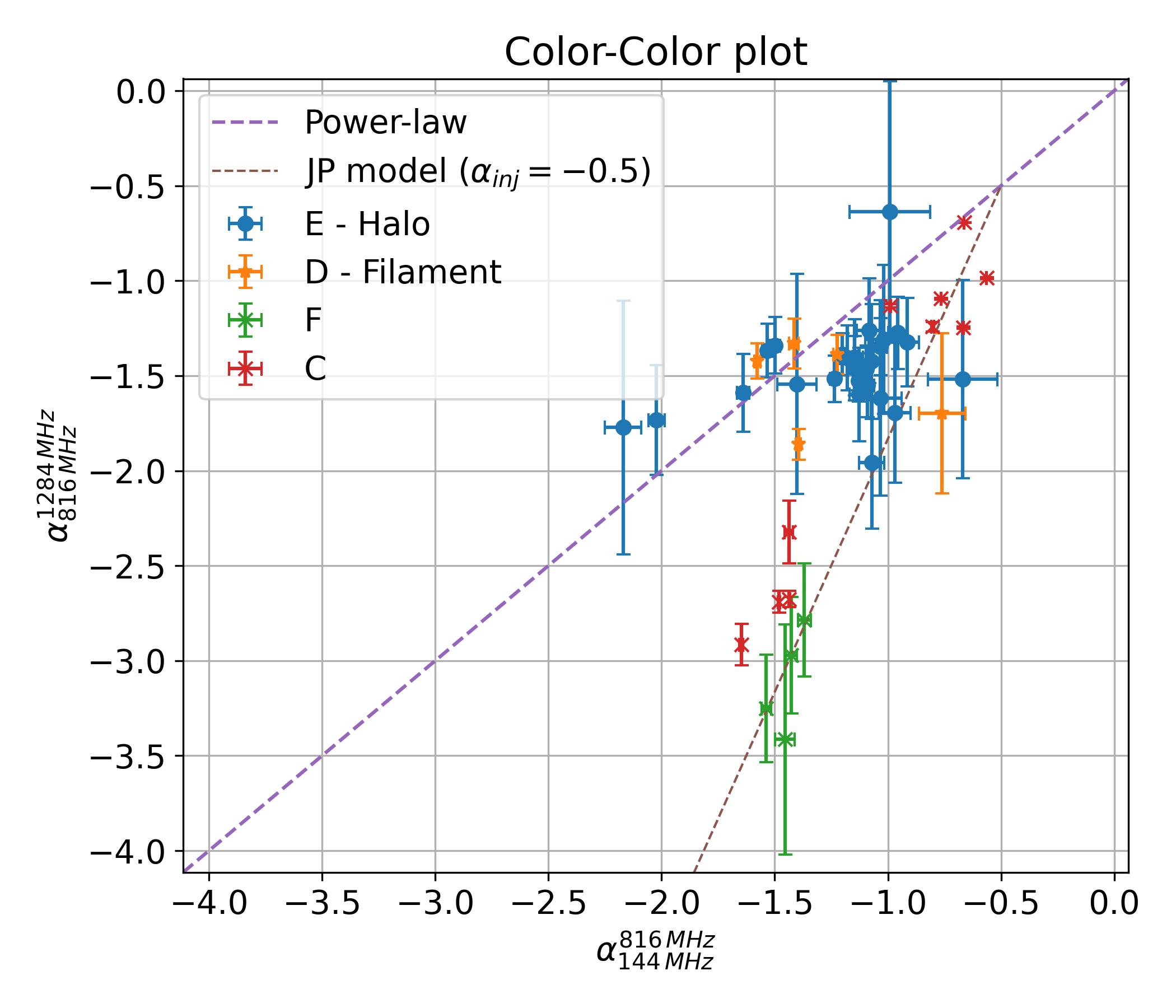}
\caption{Left panel: The L-band emission of the cluster at low resolution, shown in colors and contours. Contours start at 3$\sigma$ and are spaced by a factor of 2. The yellow circles indicate the sources that have been masked to compute the halo flux density. The bigger yellow circle is centered on the cluster X-ray peak and has a radius of 500 kpc. The cyan regions (circle and ellipse) refer to the regions E and D, as defined in \citet{Shimwell16}. The source $D_A$ in \citet{Shimwell16} is also labeled. Right panel: The color-color plot for the sources C,D,E, and F. The purple dashed line indicates the region where $\alpha_{144 MHz}^{816 MHz}= \alpha_{1284 MHz}^{816 MHz}$, i.e. a power law. The brown dashed line indicates the track of a JP model with an injection spectral index -0.5}
\label{fig:halo}
\end{figure*}

Our UHF and L-band images confirm the presence of the diffuse source E at the center of the cluster (see Figs.~\ref{fig:A2034_X-radio} and~\ref{fig:halo}). On the other hand, the eastern part of the filament detected in region D in LOFAR is not clearly distinguishable from the diffuse emission E in both 
 UHF and L-band images. We only detect a small filament of emission departing from source $D_A$, but overall the brightness of the region $D$ is similar to the brightness of the emission in the diffuse region E. 
It is likely that in the region of the filament D two different radio components are observed:  a steep spectrum and filamentary  component (Source D), which is only detected at 144 MHz, and a more diffuse component (Source E, the radio halo), which is detected in UHF and L band.
A recent analysis of the X-ray emission (Campitiello et al, submitted) indicates the presence of an X-ray discontinuity close to the source D. 
The northern part of the halo, where a shock has been detected \citep{Owers14} shows an enhancement in radio brightness, which could be linked to the shock.
\par 
Measuring the integrated flux density of the halo is challenging, due to the contamination of the other sources and of the likely superposition of the filament (source D) onto the halo (source E). 
We have masked the radio sources (see Fig.~\ref{fig:halo}) and we have measured the flux density above 3$\sigma $  within a radius of 500 kpc from the cluster center.  We have assumed that the masked areas have a brightness equal to the mean brightness of the diffuse emission. With this choice, both source E and D are part of the radio halo emission
The flux densities of the diffuse emission are listed in Table  \ref{tab:sources}. The flux density measured from the LOFAR DR2 image, in the same region, is $S_{144\, MHz} =570 \pm 60$ mJy.  \par
 In Fig.~\ref{fig:spix}, we show the spectral index map of the cluster computed between UHF and L-band images at high resolution. No trend is
  visible in the spectral index in the northern part of the halo, where a shock has been detected \citep[see][]{Owers14}, while the spectral index is steeper in the south-west part of the halo, where it reaches values of $\alpha \sim -3$.  The steep patch of emission appears to be linked to the tailed radio galaxy located to the south-west from the cluster center (source $D_A$ in \citealt{Shimwell16}, see also Fig.~\ref{fig:halo}). \par
To estimate the spectral index, we have used the LR images and computed the flux density in each frequency image in the area which is above 3$\sigma$ in the L-band image, and adopted the same approach as described above to account for the sources embedded in the diffuse emission. 
We have attempted to fit the spectrum using a single power-law $ S(\nu) \propto \nu^{\alpha}$, and obtained $\alpha= -1.43\pm  0.02$. 
The fit $\chi^2=9$ indicates that it is a poor fit of the data. Considering pairs of images, we obtain $\alpha_{144 MHz}^{816 MHz}=-1.34 \pm 0.08 $ and 
$\alpha^{1280 MHz}_{816 MHz}=-1.75 \pm 0.25$, which are not consistent within the errors and indicate a high frequency steepening of the emission, considered as a whole. A steepening of radio halos integrated spectrum has been observed already in the Coma cluster \citep[e.g.][]{Giovannini93,Bonafede22,murgia2024MNRAS.528.6470M} and interpreted in the framework of re-acceleration models \citep{BJ14}. However,
having only three frequencies, and given the challenges in accounting for the contaminating radio sources, this result should be taken with caution. \par
To further investigate the spectral properties of the sources D and E, we show in Fig. \ref{fig:halo}, right panel, the  color-color plot derived for the radio halo emission. The spectrum is computed within squared boxes having an area of seven beams, and  with a brightness larger than 3$\sigma$ in the three images. This plot shows that the radio halo is characterized by a range of spectral component ranging from -2.5 to -0.5, and confirms that several regions of the halo show a steepening at high frequency. 
The source D cannot be distinguished from the rest of the diffuse emission, likely because the diffuse emission from the halo dominates. Observations at lower frequencies (e.g. LOFAR LBA) would be required to shed light on the spectral properties of the source D. \par
Estimating the L-band power of the radio halo requires an estimate of the radio halo spectral index, which is not trivial to obtain for the reasons explained above. We have computed the power of the radio halo at 1280 MHz assuming a spectral index $\alpha = -1.75 \pm 0.25 $. The errors on the power are dominated by the uncertainty on $\alpha$. 
    We obtain $P_{1.28 \rm{GHz}}= 8.2 \pm 0.2 \cdot 10^{23}$ W/Hz. Extrapolated to 1.4 GHz,  the radio halo power agrees within the scatter with the $P_{1.4\, GHz}- M_{500}$ relation computed at 1.4~GHz \citep[e.g.][]{Cuciti21}.

\section{Magnetic field constraints from RM }
\label{sec:rm}
A scaling relation between the mass $M_{500}$ and the magnetic field within $R_{500}$ is expected from  cosmological simulations and from scaling relations of radio halos \citep{Dolag02,Balboni2025}. Specifically, it is expected that $B\propto M_{500}^{b}$, which yields for A2034 a magnetic field of 2.1 - 2.4\;$\upmu$G \citep[see][for further details]{Balboni2025}.
Here, we check whether such a magnetic field is consistent with the polarization data that we have for A2034. Indeed, the Faraday depth of sources observed through the ICM depends on the cluster magnetic field component along the line of sight, $B_{||}$ and on the gas density of the cluster $n_e$, according to:
\begin{equation}
\phi= k \int_{los} B_{||}n_e dl ,
\end{equation}
where $k$ is a constant, and the integral is along the line of sight.
In the case of an external Faraday screen, $\phi$ coincides with the Faraday Rotation Measure (RM) \citep{Brentjens05}.
In Fig.~\ref{fig:RM}, the RM image of the cluster is shown. We have divided the RM image in annuli having a width $\Delta R= \frac{1}{5} R_{500}$, and we have computed the average $\langle RM \rangle$ and the median absolute deviation $ \langle \rm{\sigma_{MAD}} \rangle $ within each annulus. 
The errors on the $\langle RM \rangle$ are computed taking into account (i) the error of the RM detection, 
(ii) the statistical error $ \rm{Err}_{\langle RM \rangle} = \frac{\sigma_{\rm RM}}{\sqrt(2 N_{beams})}$, and (iii) the uncertainty on the Galactic RM (see Sec. \ref{sec:pol}). 
The error on the $\sigma_{MAD}$ is: $ \rm{Err}_{\langle \rm{\sigma_{MAD}} \rangle} = 1.486\frac{\rm{\sigma_{MAD}}}{\sqrt(N_{beams})}$. In Fig.~\ref{fig:RM}, right panel, we show the radial profile of the $\langle RM \rangle$ and of the $\rm{\sigma_{MAD}}$ as a function of the distance from the cluster center normalized to $R_{500}$. The $\rm{\sigma_{MAD}}$ profile decreases from the cluster center towards the outskirts, as expected for a radially decreasing magnetic field and gas density distribution.
\subsection{Magnetic field 3D simulations}\label{sec:mag_field_sim}
 In order to model the cluster magnetic field, we have used the code {\tt MiRo'} \citep{Bonafede13,Stuardi22}. The code takes as input a 3D magnetic field model and a 3D gas density model, and derives mock RM images of the cluster, which can be compared with the observations to find the magnetic field parameters that best reproduce the observations. \par
 We have used the X-ray observations and derived the deprojected electron gas density radial profile $n_e(r)$ of the cluster (see Sec. \ref{sec:Xray_obs}). To account for the scatter in the density profile at each radius, we have generated a cube for the gas density profile, where each pixel contains a variable randomly extracted from a Gaussian distribution having as a mean value the value of the gas density profile at that radius ($\upmu(r) = n_e(r)$) and as a standard deviation $\sigma_G $, the standard deviation of the gas density profile at that radius ($\sigma_G  = err_{n_e}$). Using this method, the gas density fluctuations, which impact $\sigma_{\rm RM}$, are taken into account.\\
\indent  Once a model for the gas density has been built, we modeled the magnetic field assuming a power spectrum as derived in MHD cosmological simulations \citep{Dominguez-Fernandez19}. Specifically, we chose the model derived for the simulated cluster  ``E5A" which has a similar mass and dynamical status as A2034.
 The magnetic field power spectrum peaks at 230 kpc, and has components from 550 to 4 kpc, which is below the resolution of the observations. We have assumed that the magnetic field strength is proportional to the gas density as $B(r) \propto n_e^{\eta}$ and we have fixed $\eta=0.5$, as derived in the Coma cluster \citep{Bonafede10} and assumed in other works \citep{Govoni06,Vacca10,Pagliotta25}. This assumption is  physically motivated as $\eta=0.5$ corresponds to a magnetic field whose energy density is proportional to the thermal gas energy density. However, our approach is not fully self-consistent, as the scaling between the magnetic field and the gas density in the simulated cluster  ``E5A" is steeper than 0.5 (Cocchi et al, in prep.).
 The mean magnetic field is normalized within $R_{500}$ ($B_{500}$). We note that other works perform a different normalization,  \citep[e.g.][]{Murgia2004,Bonafede10,Vacca12,Osinga25}, which assumes a value $B_0$ at the cluster center and enforce a radial scaling $B \propto B_0 n^{\eta}$, though a normalization over a larger volume has already been used by \citet{Stuardi21,Pagliotta25}. We chose to normalize over $R_{500}$ and propose this approach to better compare the magnetic fields estimates in clusters with different dynamical status and large asymmetries in the radial profile. \\
\indent  We have explored a range of $B_{500}$, including the values $B_{500}= 2.1-2.4\, \upmu$G, as predicted by the scaling relations mentioned above.
 The values we have tested are listed in  the Table \ref{tab:Bsim}.
 For each magnetic field model, we have realized 30 simulations to account for the statistical scatter of each magnetic field model.
 Once the mock RM image has been obtained, for each simulation, it has been convolved to a Gaussian function having the same standard deviation as the beam of the RM observations. Smoothing the simulated RM map mimics the effect that a finite resolution has of $\sigma_{MAD}$ and $\langle RM \rangle$. To have a fair comparison with observations, each RM mock image has been masked as the RM image of the cluster, i.e. we consider from the simulations only the regions where we have recovered RM in the observations.
In addition, to account for possible bandwidth depolarization, we have masked the pixels in the simulated RM image with a value larger than $|\phi_{max}|= 2000 \, \rm{rad/m^2}$ (see Eq. \ref{eq:phimax}).\par
\begin{figure*}
\includegraphics[width=\columnwidth]{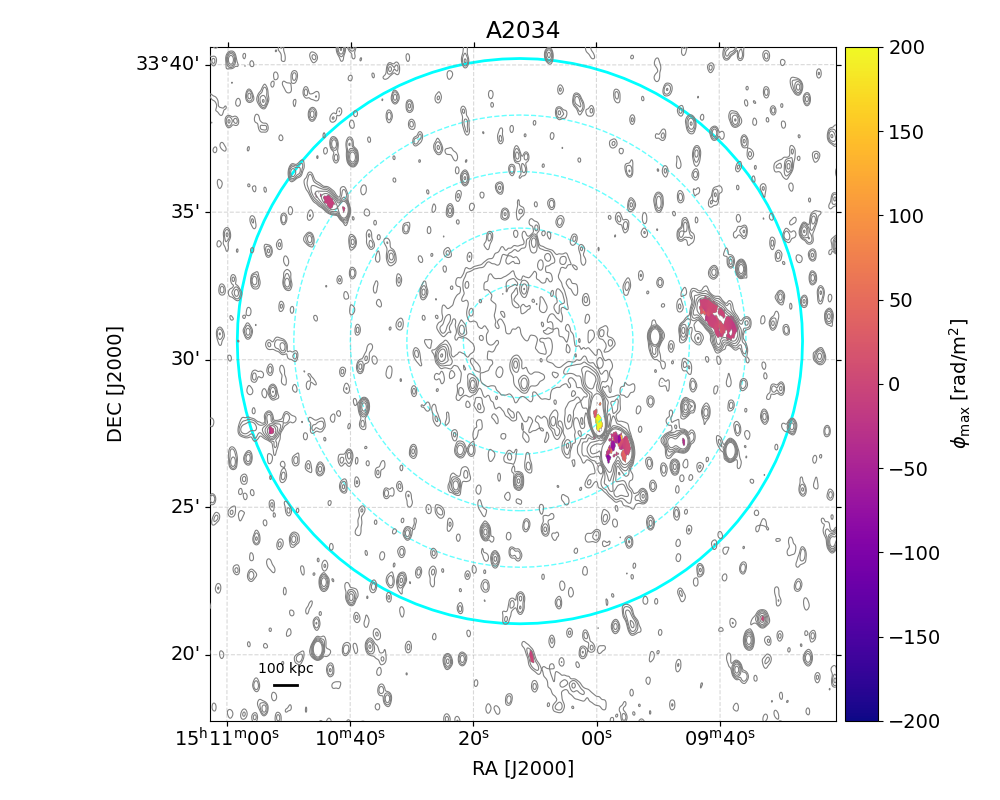}
\includegraphics[width=\columnwidth]{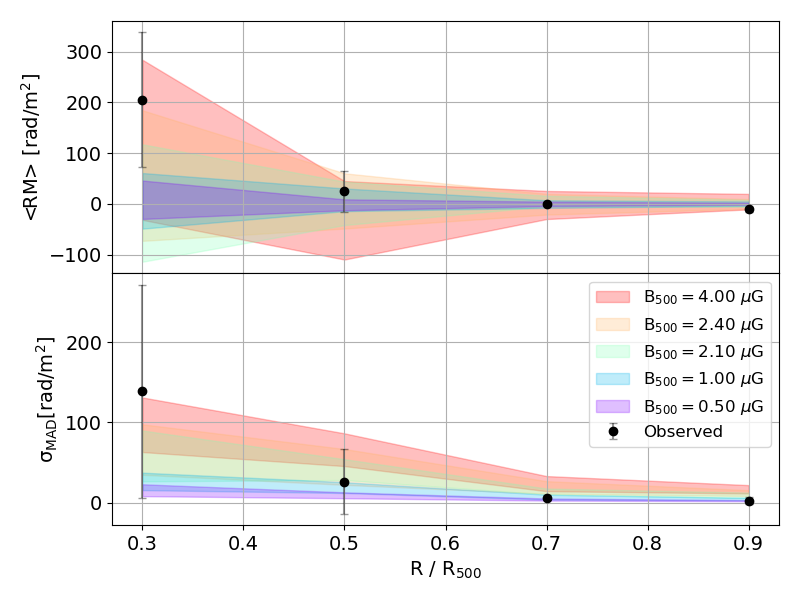}

\caption{RM image and radial profile. Left panel the RM image of the cluster is shown in colors, while contours refer to the L-band emission at the resolution of 18\arcsec $\times$ 10\arcsec. Contours start at 3$\sigma$ (1$\sigma=9 \, \upmu$Jy/beam) and are spaced by a factor of 2. The cyan circle marks $R_{500}$ and the cyan dashed annuli are spaced by 1/5 $R_{500}$ and represent the regions used for the average profile of $\langle RM \rangle$ and $\rm{\sigma_{MAD}}$. Right panel: simulated and observed profiles of $\langle RM \rangle$ (top) and $\rm{\sigma_{MAD}}$ (bottom) versus the cluster radius, normalized to $R_{500}$. }
\label{fig:RM}
\end{figure*}

 From the 30 different simulations, we have derived the distribution of RM means and $\rm{\sigma_{MAD}}$, which we have compared with the corresponding observed quantities. The observed and simulated profiles of $\langle RM \rangle$ and $\rm{\sigma_{MAD}}$ are shown in Figure \ref{fig:RM}. The reduced $\chi^2$ for each simulation is listed in Table \ref{tab:Bsim}.\\
\indent From the comparison, we conclude that
no value of $B_{500}$ provides a  fit to the data in both $\sigma_{MAD}$ and $\langle RM \rangle$ with $\chi^2_r \leq 1$. In particular, while the $\langle RM \rangle$ values are compatible with magnetic fields $B_{500} \geq 1.0 \, \upmu$G, none of the model we have considered provides a $\chi^2_r \leq 1$ to the $\sigma_{MAD}$. The best agreement of the $\sigma_{MAD}$ trend is obtained for  $B_{500} \leq 1.0 \, \upmu$G. The $\chi_r^2$ fit is driven by the points at large distance from the cluster center, which have a small $\sigma_{MAD}$. A residual intrinsic RM and/or a non-perfect subtraction of the Galactic emission could contribute to the observed values, and determine the high value of the $\chi_r^2$. We also note that we miss constraints at $R< 0.3 \, R_{500}$, where models would show the largest differences. Indeed, Fig.~\ref{fig:RM} indicates that reducing the uncertainty on the inner point would be crucial to discriminate the models.\\
\indent 
Deriving the best-fit magnetic field is beyond the scope of this work, since the limited observational constraints require several assumptions on the magnetic field model to proceed with the comparison between data and simulations. 
In particular, our procedure does not fully account for the changes in the distribution of simulated  RMs that could occur due to either beam or line of sight depolarization. Hence, we stress that our results depend on the assumptions on the gas density and magnetic field model that we have made, under which a values of $B_{500} = 1\upmu$G provides the best fit to the data.

\begin{table}
\centering
\begin{tabular}{ccc}
\hline\hline 
 $B_{\rm 500}$ &  $ \chi_r^2$ on ${\rm \sigma_{MAD}}$&   $\chi_r^2$ on ${\rm \langle RM \rangle }$ \\
\;$\upmu$G  & & \\
 \hline
0.5  &1.8 & 2.6 \\
1.0  & 1.8 & 0.9 \\
2.1  & 2.9 & 1.3 \\
2.4  & 3.3 & 0.6 \\
4.0  & 4.7  & 0.5 \\
\hline \hline
 
\end{tabular}
\caption{Magnetic field simulations: reduced $\chi^2$ values for different $B_{500}$}
\label{tab:Bsim}
\end{table}

\section{Discussion and Conclusions}
\label{sec:discussion}
\label{sec:conclusions}
In this work, we have used new MeerKAT data in the  UHF and L-band to derive the spectral and polarimetric properties of the galaxy cluster A2034. 
The diffuse sources discovered by previous studies \citep{Giovannini09, vanWeeren11,Shimwell16} were of uncertain classification. In particular, LOFAR observations had revealed a plethora of diffuse emission, including an irregular radio halo, three candidate radio relics, and steep-spectrum filaments not clearly connected to cluster members or AGN. \par
Our results are can be summarized as follows:
\begin{itemize}
\item We confirm the presence of diffuse emission at the cluster center, with irregular morphology and no detectable polarization ($F_P < 0.1 $). The integrated spectrum measured within a circle of 500 kpc radius steepens from $\alpha_{144 MHz}^{816 MHz}=-1.34 \pm 0.08 $  to $\alpha_{816 MHz}^{1.28 GHz}=-1.75 \pm 0.25 $. 
The spectral index distribution is not uniform, and in particular it steepens towards the south, in the direction of the source $D_A$. The filament D  is only partially detected in UHF and L band with respect to LOFAR. A resolved analysis of the spectrum of the source E and D  (Fig. \ref{fig:halo}) reveals that several regions of source E show a steepening at high frequency, while other regions follow a power law with $\alpha$ ranging from -0.5 to -2.5.  The power of the radio halo follows the $P_{\rm 1.4 \, GHz} - M_{500}$ relation found in the literature \citep[e.g.][]{Cuciti21}, despite its unusually steep spectrum.
\item The northern part of the radio halo is coincident with a shock detected in the X-rays. The radio emission is brighter in this region, but no sharp radio edge nor spectral index gradient is detected. Hence, we conclude that the radio emission is unlikely to be powered by shock (re)-acceleration as in the case of radio relics.
\item Source A in \cite{Shimwell16} presents typical features of radio relics, such as average fractional polarization of $F_p \sim 7\%$, and polarization vectors that trace a magnetic field aligned with the main axis of the source. These features are in agreement with the expectations from shock (re)-acceleration, though ad-hoc projection effects should be assumed to explain the inverted trend in the spectral index distribution detected along the relic main axis.
Assuming standard DSA conditions, we have estimated from the radio spectral index a Mach number $M =1.59 \pm 0.07$ which would power the relic emission.
\item The source F, located 1380 kpc to the south of the cluster center exhibits a steep spectrum ($\alpha_{144 MHz}^{816 MHz} = -1.5 \pm 0.2$)  which makes it barely visible in the MeerKAT L-band at high resolution at our sensitivity threshold. A resolved spectral index analysis using the color-color plot (Fig. \ref{fig:halo}) suggests that the source spectrum is consistent with a JP model. 
The patches of emission detected in the L-band appear polarized at low resolution, though because of Galactic contamination, we cannot unambiguously determine the fractional polarization of the source. No spectral index gradient is detected along the main axis of the source. Overall, source F does not have the features expected for radio relics, and we suggest that it is aged emission from a radio galaxy, though the morphology suggests a compression/interaction with the ambient medium.
\item Using the Faraday depth radial profile of the sources observed through the ICM of A2034, and modeling the gas density component from X-ray observations, we have investigated the properties of the ICM magnetic field.
In particular, our aim was to check whether the magnetic field expected for a cluster as massive as A2034 would reproduce a $\langle RM \rangle$ and $\rm{ \sigma_{MAD}}$ profile as observed.
 We have used X-ray observations to derive the cluster gas density profile, and we have taken into account the radial scatter of the gas density to derive mock RM images of the cluster with different normalization for the magnetic field. 
Because of the low number of detected sources in polarization, we have assumed a magnetic field power spectrum as derived in MHD simulations \citep{Dominguez-Fernandez19} and a magnetic field dependence on the gas density profile: $B(r) \propto n_e^{0.5}$.
Under these assumptions, a magnetic field $B_{500} = 1\upmu $G  provides the best fit to  $\rm{ \sigma_{MAD}}$ and $\langle RM \rangle$ radial profile, though the detection of polarized sources at $r< 0.3 R_{500}$ would be critical to discriminate among different magnetic field normalizations.
\end{itemize}
Our results can  be discussed in the framework of the merger geometry proposed for the system. \citet{Owers14} concluded that X-ray and optical data are  consistent with a merger of two main sub-clusters, which are  now moving apart along a north-south axis after a small impact parameter core passage. 
The presence of the steep filament D, close to an X-ray edge detected by a recent work (Campitiello et al. submitted), could be related to the sub-cluster moving toward the south. On the other hand, the presence of a relic at the east of the cluster (source A), under the accepted scenario that relics are powered by shocks, indicates a more complex merger, possibly involving more than two sub-clusters. \par
The analysis presented in this work underlines the importance of multi-frequency deep and polarimetric observations to understand the origin of the diffuse emission detected in galaxy clusters, and the need for a large sample of sources detected in polarization to constrain the magnetic field in the ICM.


\bibliographystyle{aa} 
\bibliography{biblio}

\begin{acknowledgements}
AB and MBalboni acknowledge support from the ERC CoG $\vec{B}$ELOVED, GA N.101169773. The MeerKAT telescope is operated by the South African Radio Astronomy Observatory, which is a facility of the National Research Foundation, an agency of the Department of Science and Innovation.
LOFAR is the Low Frequency Array designed and constructed by ASTRON. It has observing, data processing, and data storage facilities in several countries, which are owned by various parties (each with their own funding sources), and which are collectively operated by the LOFAR ERIC under a joint scientific policy. The LOFAR resources have benefited from the following recent major funding sources: CNRS-INSU, Observatoire de Paris and Université d'Orléans, France; BMBF, MIWF-NRW, MPG, Germany; Science Foundation Ireland (SFI), Department of Business, Enterprise and Innovation (DBEI), Ireland; NWO, The Netherlands; The Science and Technology Facilities Council, UK; Ministry of Science and Higher Education, Poland; The Istituto Nazionale di Astrofisica (INAF), Italy.
MBr\"u acknowledges funding by the Deutsche Forschungsgemeinschaft (DFG) under Germany's Excellence Strategy -- EXC 2121 ``Quantum Universe" --  390833306 and the DFG Research Group "Relativistic Jets". 
CJR acknowledges financial support from the DFG, via the Collaborative Research Center SFB1491 `Cosmic Interacting Matters – From Source to Signal'.
FdG acknowledges support from the ERC Consolidator Grant ULU 101086378.
This work made use of Astropy:\footnote{http://www.astropy.org} a community-developed core Python package and an ecosystem of tools and resources for astronomy.
Part of the data published here have been reduced using the CARACal pipeline, partially supported by ERC Starting grant number 679627 “FORNAX”, MAECI Grant Number ZA18GR02, DST-NRF Grant Number 113121 as part of the ISARP Joint Research Scheme, and BMBF project 05A17PC2 for D-MeerKAT. Information about CARACal can be obtained online under the URL: https://caracal.readthedocs.io”.
 MBri acknowledges the financial contribution from the INAF GO grant 1.05.24.02.10 Extended Radio Emission in Galaxy Clusters at deep focus with MeerKAT.
\end{acknowledgements}

\begin{appendix}
\onecolumn
\section{Check of flux scales}
\label{appendix:fluxscale}
\begin{figure}
\centering
\includegraphics[width=0.45\textwidth]{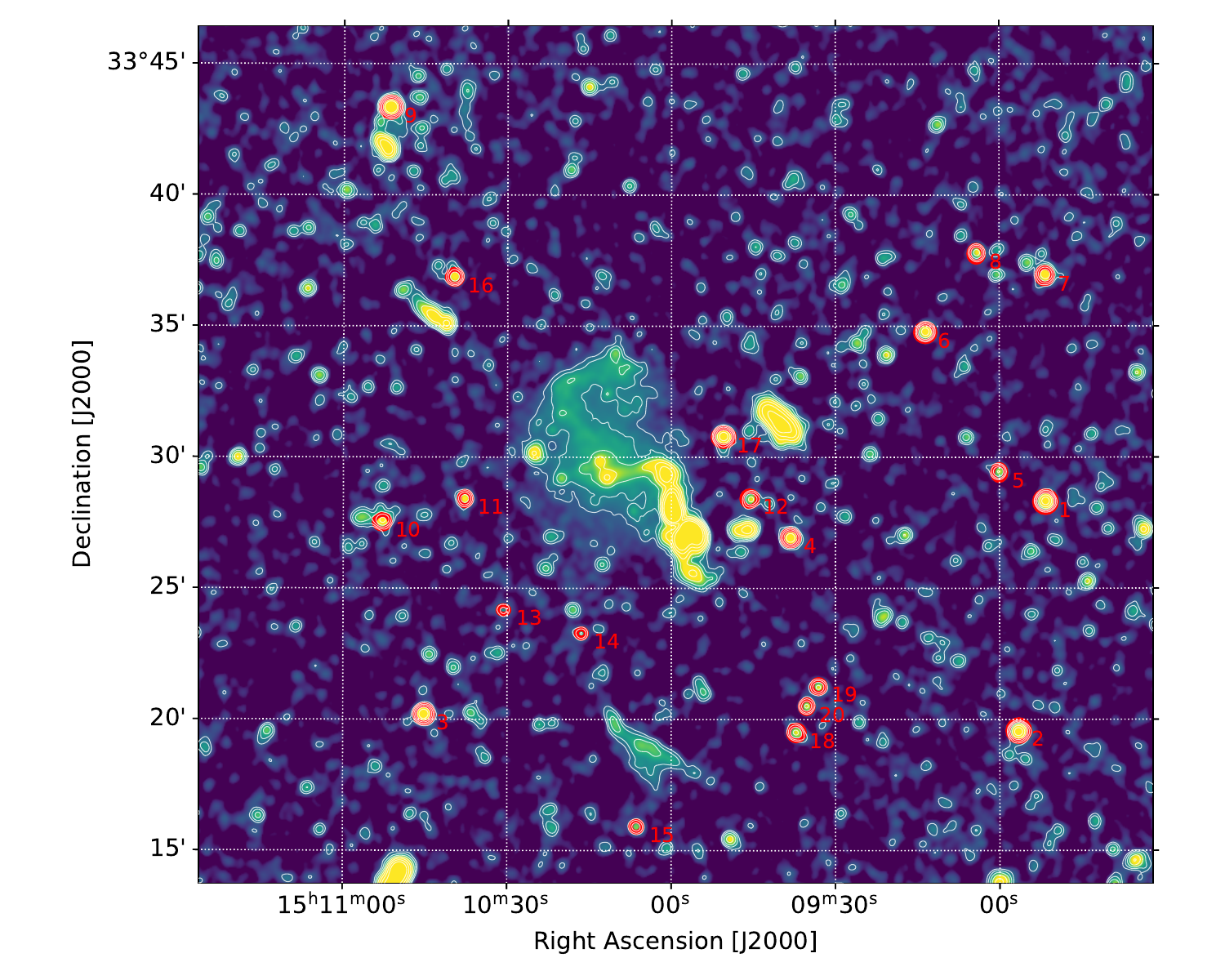}
\includegraphics[width=0.45\textwidth]{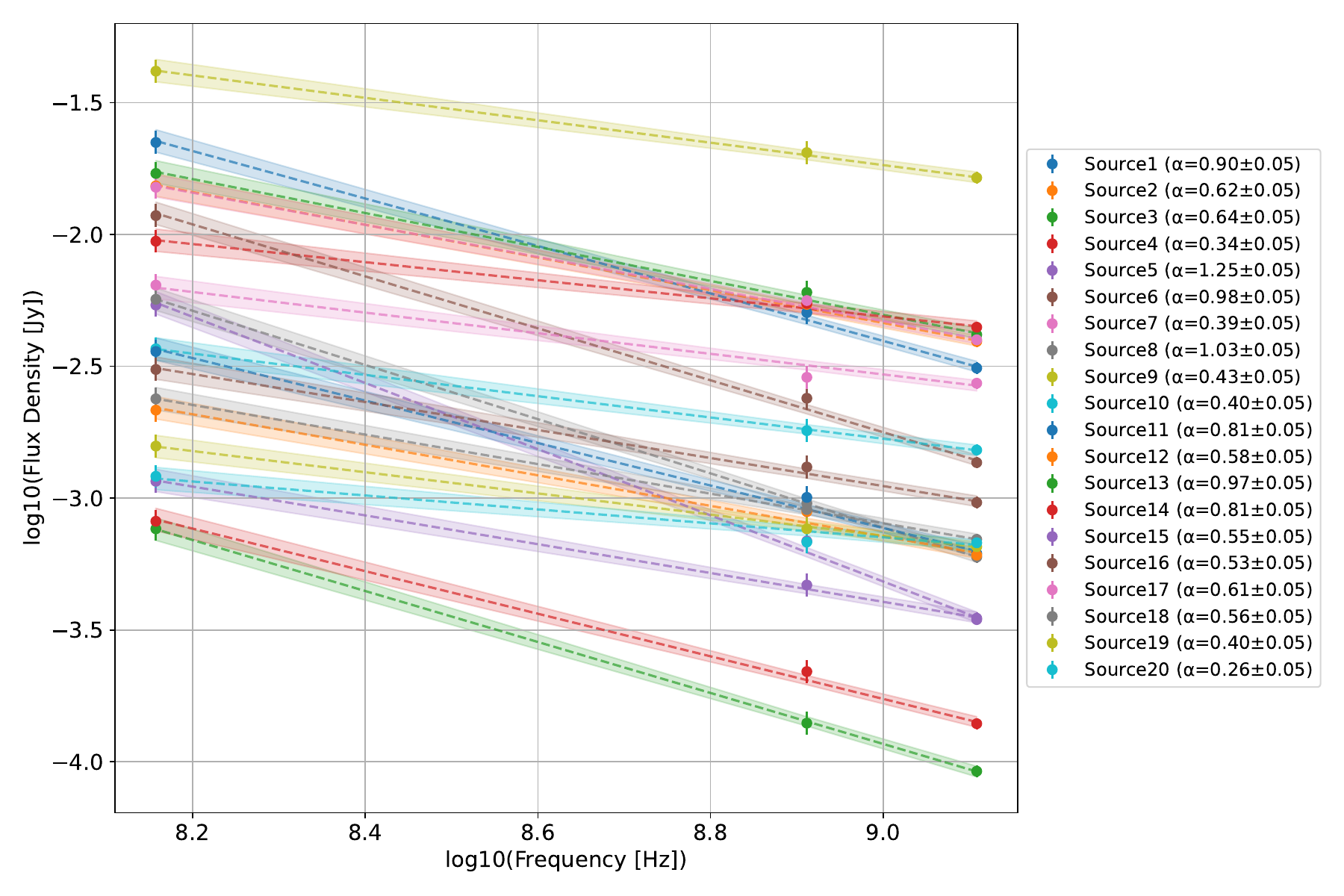}
\caption{Sources used to check the flux scale. Left panel: UHF image at low resolution in colors. Contours are from the same image, they start at 5$\sigma$ and spaced by a factor of 2. Circles and numbers label the sources used to check the flux scale alignment and correspond to the spectra shown in the right panel. Right panel: spectra of unresolved sources in the Abell~2034 field. The derived spectral index $\alpha$ and its 1$\sigma$ uncertainty ($\alpha_{err}$) are reported. The dashed lines show the best fit relations, while the shaded areas represents the $1\sigma$ confidence interval.}
\label{fig:Sources_check}
\end{figure}

In order to verify that the steepening of the emission detected between UHF and L-band images is not caused by systematics, such as different flux scales and re-normalization performed on
 the LOFAR DR2 images, we have computed the spectrum of unresolved sources in the field. Specifically, we have selected all the unresolved sources in the field of view of A2034 that appear detected at more than 3$\sigma$ in all the images (with the cut driven by the LOFAR image). The resulting twenty sources are marked by red circles in Fig.~\ref{fig:Sources_check}. 
We have computed the flux density of these sources from the LOFAR image, as well as from the MeerKAT UHF and L-band image, at low resolution, ad we have fitted their spectrum assuming a single power-law. 
We list in Table \ref{tab:appendix_sources} the derived spectral index, with relative uncertainty and the $\chi^2$ value of the fit. We show the fits in the right panel of Fig.~\ref{fig:Sources_check}. The reduced $\chi^2$ value indicates for all but three sources a good fit, indicating that the spectrum of the sources can be described by a single power-law, with an average value of $\alpha = -0.65$. From the plot of the spectra of the sources, we do not detect any systematic trend that could point towards a misalignment of the flux scale.

\begin{table}
\caption{Flux density of unresolved sources in Abell~2034 field }
\centering
\begin{threeparttable}
\begin{tabular}{lcc}
\hline\hline
Name  &  $\alpha  \pm \alpha_{err} $ &  Reduced $\chi^2$ \\
Source 1 & $\alpha$ = 0.900  $\pm$ 0.05 &   0.49  \\
Source 2 & $\alpha$ = 0.620  $\pm$ 0.05 &  0.15   \\
Source 3 & $\alpha$ = 0.643  $\pm$ 0.05 &  0.51    \\
Source 4 & $\alpha$ = 0.343  $\pm$ 0.05 &   0.23  \\
Source 5 & $\alpha$ = 1.254  $\pm$ 0.05 &   1.14  \\
Source 6 & $\alpha$ = 0.985  $\pm$ 0.05 &   1.10  \\
Source 7 & $\alpha$ = 0.391  $\pm$ 0.05 &   1.31  \\
Source 8 & $\alpha$ = 1.028  $\pm$ 0.05 & 0.001    \\
Source 9 & $\alpha$ = 0.425  $\pm$ 0.05 & 0.061    \\
Source 10 & $\alpha$ = 0.404  $\pm$ 0.05 &   0.01 \\
Source 11 & $\alpha$ = 0.805  $\pm$ 0.05 &   1.27 \\
Source 12 & $\alpha$ = 0.579  $\pm$ 0.05 &   1.15 \\
Source 13 & $\alpha$ = 0.966  $\pm$ 0.05 & 0.02   \\
Source 14 & $\alpha$ = 0.807  $\pm$ 0.05 &   0.66\\
Source 15 & $\alpha$ = 0.546  $\pm$ 0.05 &  0.14  \\
Source 16 & $\alpha$ = 0.531  $\pm$ 0.05 &   0.40 \\
Source 17 & $\alpha$ = 0.609  $\pm$ 0.05 &   0.37 \\
Source 18 & $\alpha$ = 0.560  $\pm$ 0.05 &  0.004  \\
Source 19 & $\alpha$ = 0.401  $\pm$ 0.05 &  0.05  \\
Source 20 & $\alpha$ = 0.265  $\pm$ 0.05 & 1.07   \\

\hline\hline 
\end{tabular}
\begin{tablenotes}
\item  
\end{tablenotes}
\end{threeparttable}
\label{tab:appendix_sources}

\end{table}


\section{The Spectral index error maps}
\label{appendix:spixerror}
In this Appendix, we show the errors on the spectral index images shown in the main text.

\begin{figure}
\centering
\includegraphics[width=0.5\columnwidth]{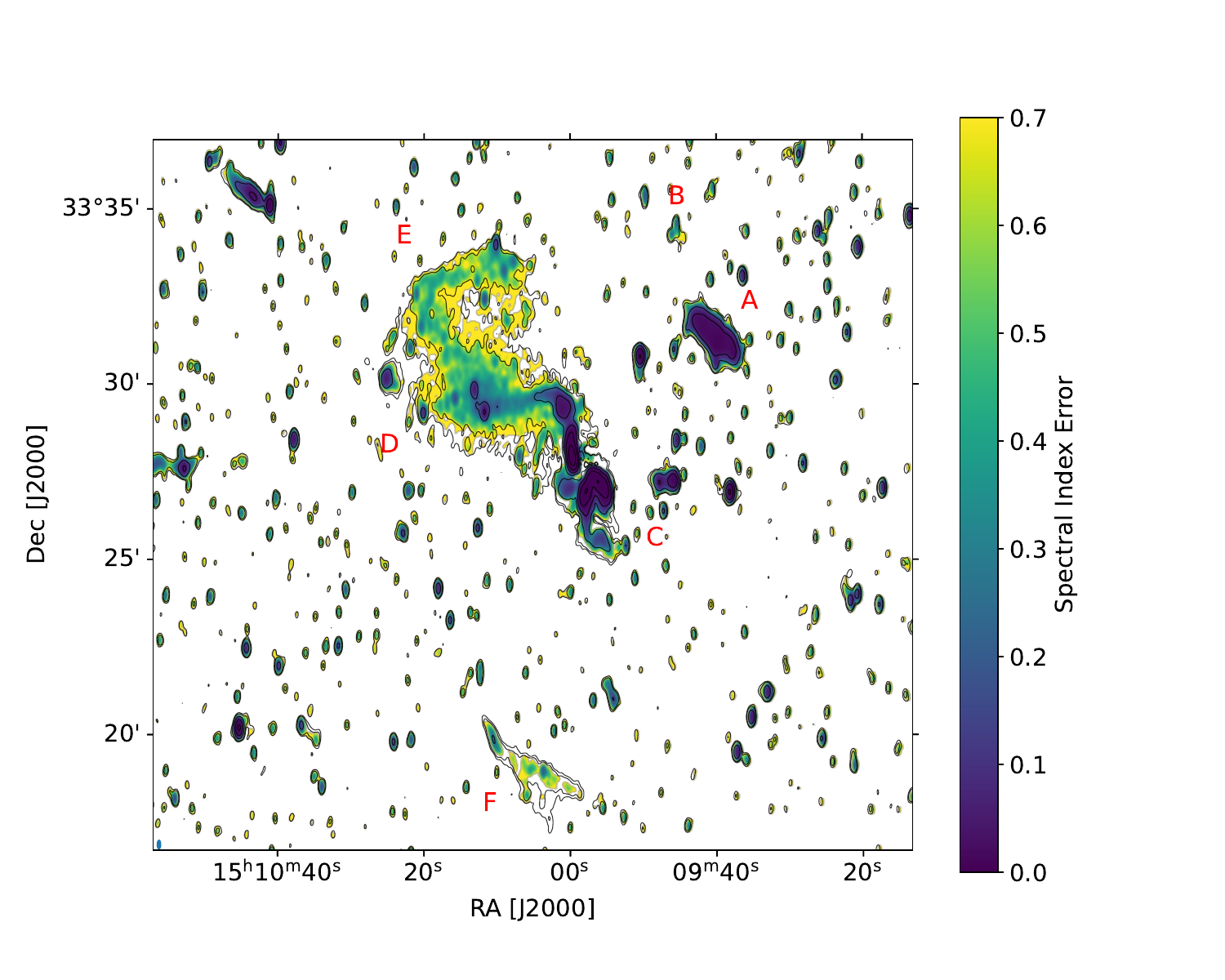}
\caption{Spectral index error image of the cluster Abell~2034 computed between the L-band and UHF Band at the resolution of $18\arcsec \times 10 \arcsec$ in colors. Pixels below 3$\sigma$ in both images have been blanked. Contours refer to the UHF band at the same resolution, contours are drawn at 3$\sigma$,5$\sigma$ and are then spaced by a factor of 4. The rms noise $\sigma$ is 15\;$\upmu$Jy/beam. The beam HPBW is $18 \arcsec \times 8 \arcsec$. Labels refer to the sources identified in \citealt{Shimwell16}. The corresponding spectral index map is shown in Fig.~\ref{fig:spix}.}
\label{fig:spix_appendix}    
\end{figure}

\begin{figure}
\centering
\includegraphics[width=0.5\columnwidth]{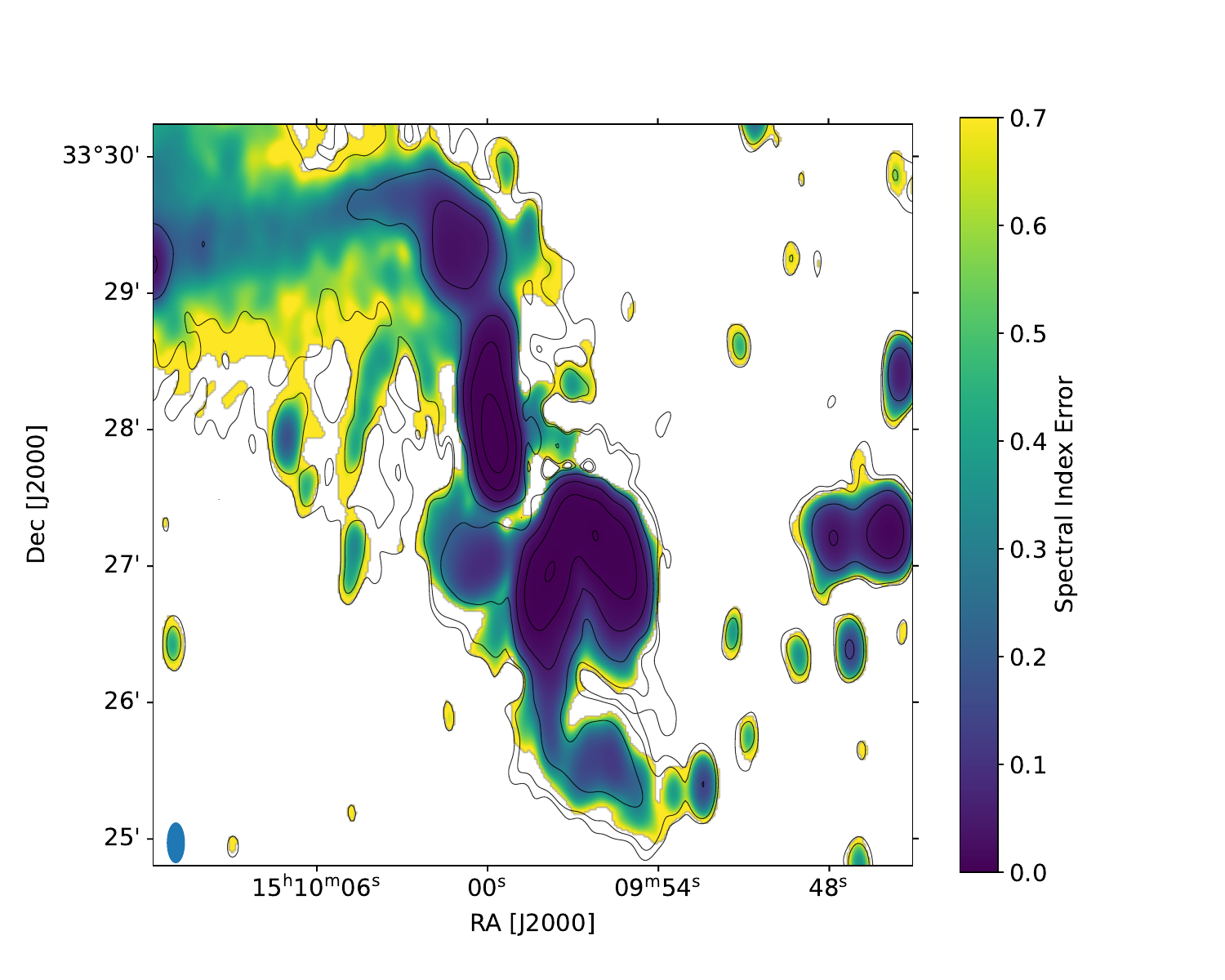}
\caption{The complex region C: Spectral index error image computed between UHF band and L-band at high resolution (HR images). Contours show the emission from the UHF image. Contours start at 3$\sigma$ and are spaced by a factor of 2. The corresponding spectral index map is shown in Fig.~\ref{fig:sourceC}.}
\label{fig:sourceC_appendix}
\end{figure}

\section{The Galactic Faraday Depth}
\label{appendix:GalacticRM}
\begin{figure}
\includegraphics[width=0.45\textwidth]{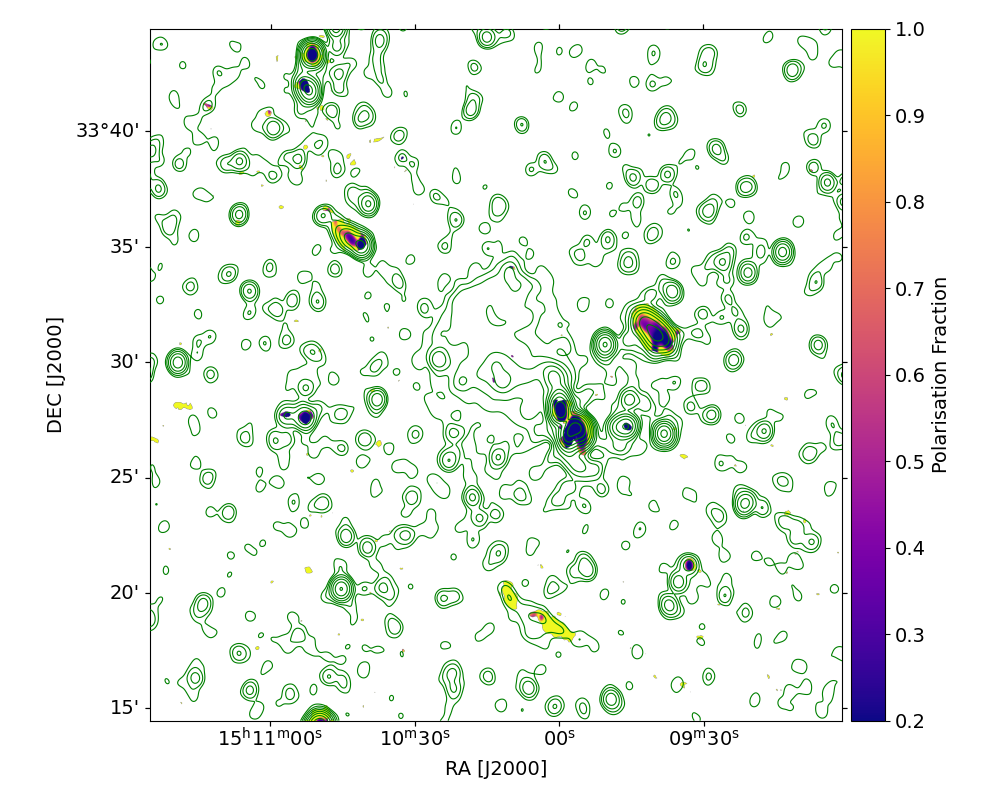}
\centering
\caption{Apparent fractional polarization from the field of Abell~2034, obtained imposing a threshold on $F_{\phi}$ only and not on I. The contours refer to the total intensity emission in L-band. The first contour is drawn at 2$\sigma$, and following contours are spaced by a factor of 2.}
\label{fig:A2034_FP_noCut}
\end{figure}

\begin{figure}
\includegraphics[width=0.45\columnwidth]{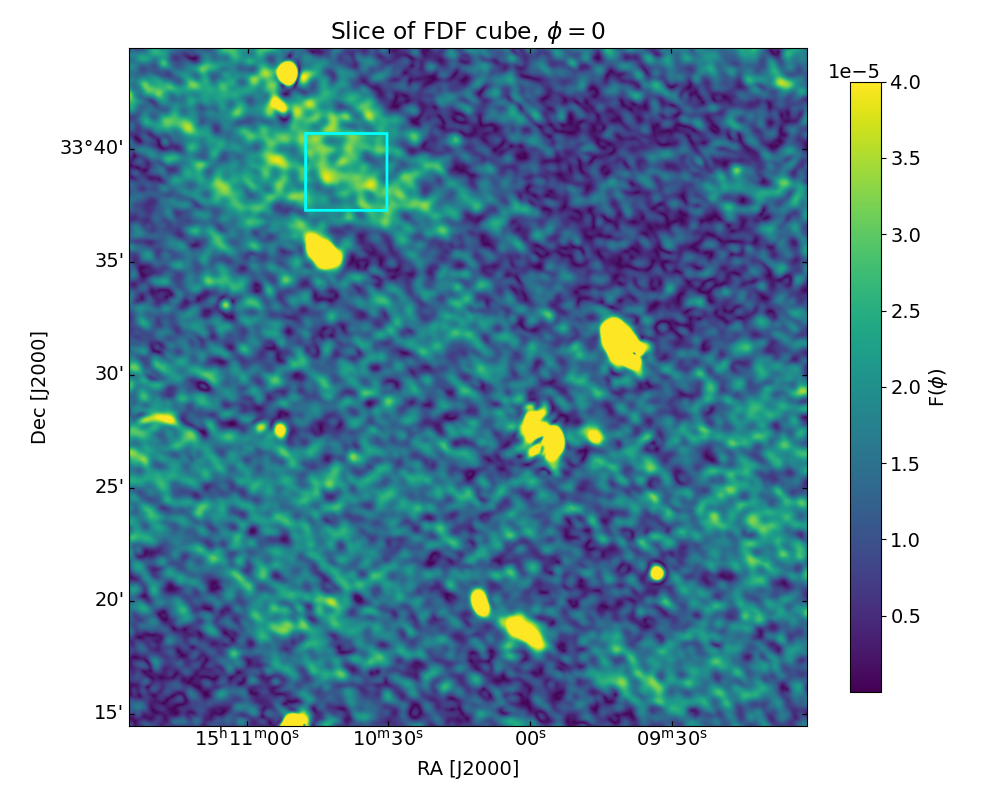}
\includegraphics[width=0.45\columnwidth]{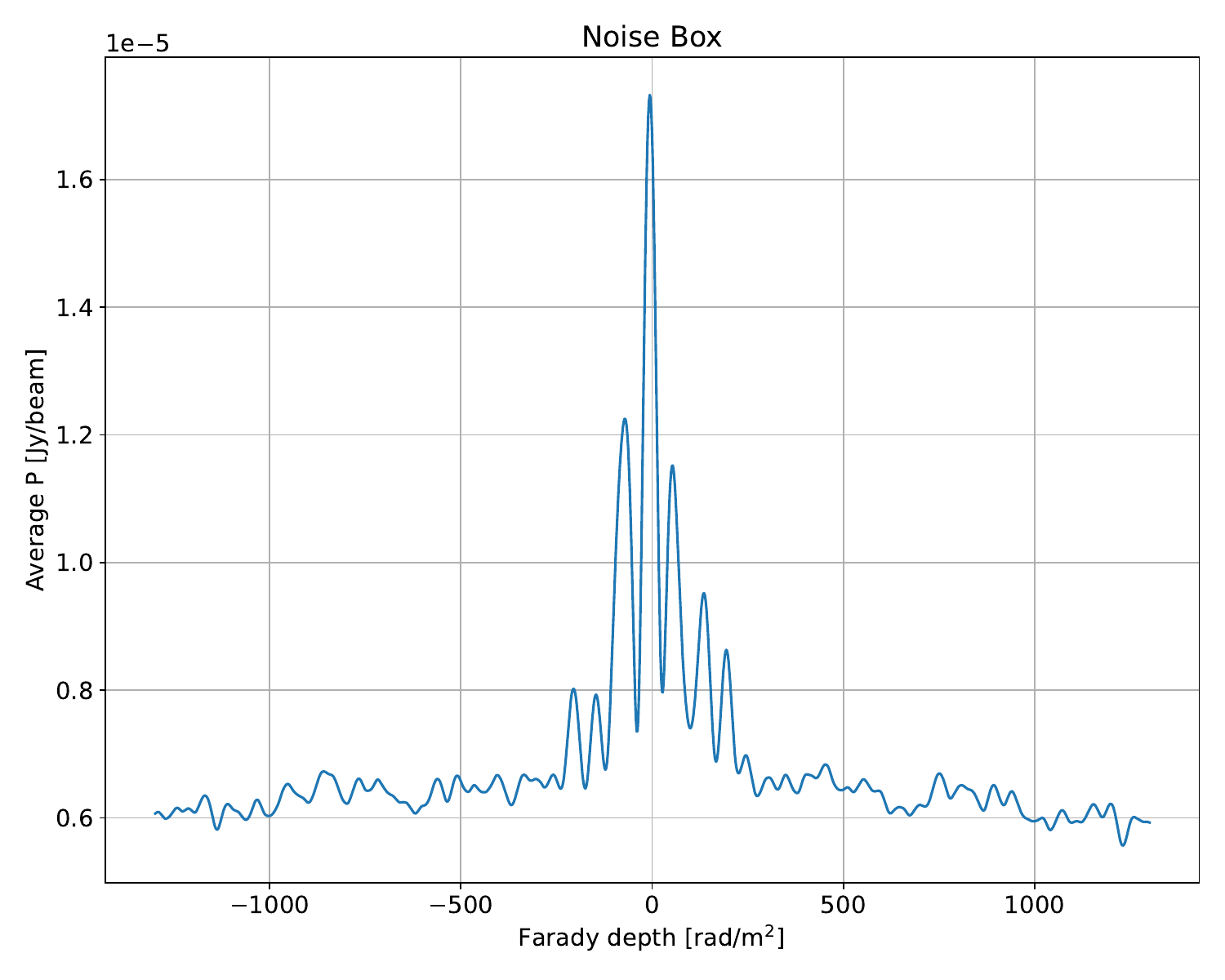}
\includegraphics[width=0.45\columnwidth]{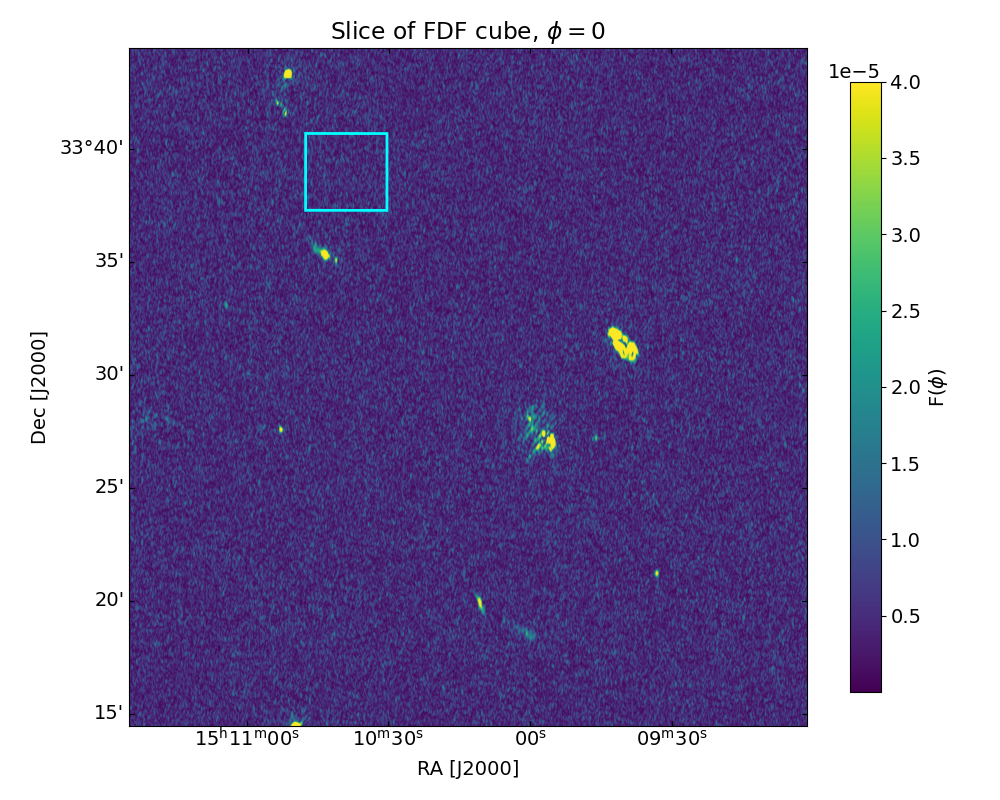}
\includegraphics[width=0.45\columnwidth]{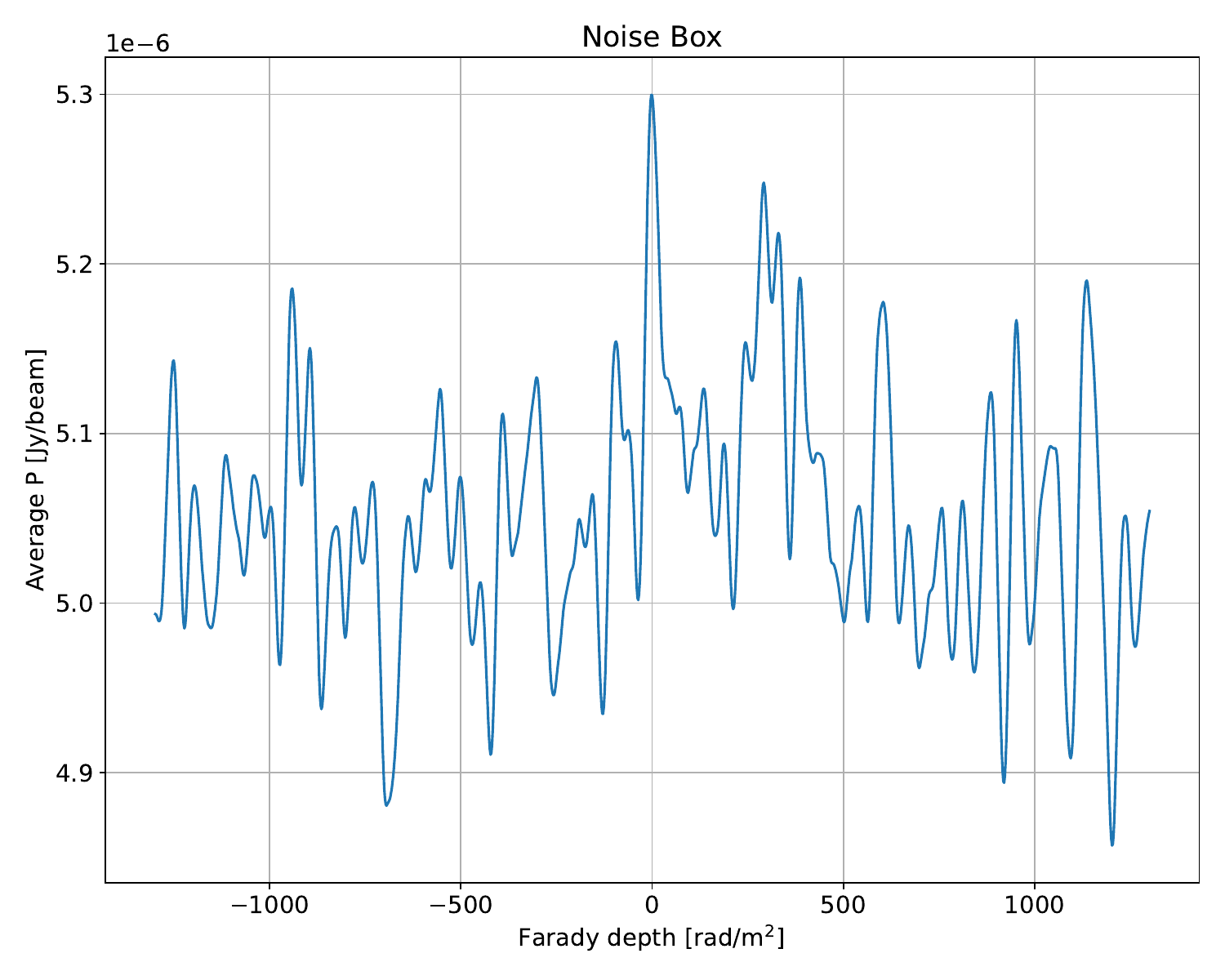}
\caption{Galatic polarized emission. Top left panel: Slice of the FDF cube centered at $\phi=0\, \rm{rad/m^2}$ obtained from the LR images in polarization. Top right panel: FDF average spectrum of the region shown in the left panel (cyan box).
Bottom left panel: same as top left panel but from the HR image. Bottom right panel: same as top right panel but from the HR image.}
\label{fig:FDF_noise}
\end{figure}

In order to search for diffuse polarized emission of the source F, we have imaged the L-band data in Q and U at low resolution, using a Briggs weighting scheme with robust $r=-0.5$, and a UV-taper of 15\arcsec. The images have been convolved to a Gaussian beam of 21\arcsec. We have processed the images using the RM-synthesis technique, as explained in Sec. \ref{sec:rm} for the high resolution images. The low resolution image displaying the peaks of $F(\phi)$ is shown in Fig.~\ref{fig:A2034_FP_noCut}. When a cut at $6 \sigma$ is imposed, we recover polarized emission from the same regions as obtained in the high resolution image, plus few small regions 
that therefore appear at very high values of fractional polarization (see Fig.~\ref{fig:A2034_FP_noCut})
However, when we average the $F(\phi)$ spectrum over large regions that do not correspond to total intensity emission, a clear peak at $\phi= 0 \rm{rad/m^2}$ is detected (see Fig \ref{fig:FDF_noise}). 
For comparison, we also show the spectrum from the same region at high resolution, where no clear peak is detected

We interpret this emission as Galactic emission, which is filtered out at high resolution. 
We show in Fig. \ref{fig:FDF_noise} a slice of the $F(\phi)$ cube, corresponding to $\phi=0 \rm{rad/m^2}$ (after removal of the Galactic RM). We see that coherent patches of emission are visible.
The Galactic emission is not strong enough to be detected on single pixels, but it may add a non negligible contribution to the value of the polarization of extended sources detected al low resolution only (e.g. source F).

\end{appendix}

\end{document}